%% file: paper.tex
\documentclass[sigconf, nonacm]{acmart}

\newcommand\vldbdoi{10.14778/3551793.3551833}
\newcommand\vldbpages{2811 - 2825}
\newcommand\vldbvolume{15}
\newcommand\vldbissue{11}
\newcommand\vldbyear{2022}

\newcommand\vldbtitle{\shorttitle} 
\newcommand\vldbavailabilityurl{}
\newcommand\vldbpagestyle{plain}

\input{header}

\newcommand{\myparagraph}[1]{\vspace{1.5mm}\noindent\textbf{#1}}

\begin{document}

\title{Query Processing on Tensor Computation Runtimes} %

\author{Dong He$^{1}$, Supun Nakandala$^{2}$, Dalitso Banda$^3$, Rathijit Sen$^3$, Karla Saur$^3$, Kwanghyun Park$^3$,\\ Carlo Curino$^3$, Jes\'us Camacho-Rodr\'iguez$^3$, Konstantinos Karanasos$^4$, Matteo Interlandi$^3$}

\affiliation{
    \institution{$^1$University of Washington, $^2$University of California, San Diego, $^3$Microsoft, $^4$Meta}
}
\affiliation{$^1$donghe@cs.washington.edu, $^2$snakanda@eng.ucsd.edu, $^3$firstname.lastname@microsoft.com, $^4$kkaranasos@fb.com}

\input{abstract}

\maketitle

\pagestyle{\vldbpagestyle}
\vspace{-1ex}
\begingroup\small\noindent\raggedright\textbf{PVLDB Reference Format:}\\
Dong He, Supun Nakandala, Dalitso Banda, Rathijit Sen, Karla Saur, Kwanghyun Park, Carlo Curino, Jes\'us Camacho-Rodr\'iguez, Konstantinos Karanasos, Matteo Interlandi. \vldbtitle. PVLDB, \vldbvolume(\vldbissue): \vldbpages, \vldbyear.\\
\href{https://doi.org/\vldbdoi}{doi:\vldbdoi}
\endgroup
\begingroup
\renewcommand\thefootnote{}\footnote{\noindent Work done while Dong, Supun, and Konstantinos were at Microsoft. 

\noindent This work is licensed under the Creative Commons BY-NC-ND 4.0 International License. Visit \url{https://creativecommons.org/licenses/by-nc-nd/4.0/} to view a copy of this license. 
For any use beyond those covered by this license, obtain permission by emailing \href{mailto:info@vldb.org}{info@vldb.org}. 
Copyright is held by the owner/author(s). Publication rights licensed to the VLDB Endowment. \\
\raggedright Proceedings of the VLDB Endowment, Vol. \vldbvolume, No. \vldbissue\ %
ISSN 2150-8097. \\
\href{https://doi.org/\vldbdoi}{doi:\vldbdoi} \\
}\addtocounter{footnote}{-1}\endgroup

\ifdefempty{\vldbavailabilityurl}{}{
\vspace{.3cm}
\begingroup\small\noindent\raggedright\textbf{PVLDB Artifact Availability:}\\
The source code, data, and/or other artifacts have been made available at \url{\vldbavailabilityurl}.
\endgroup
}

\begin{sloppypar}

\input{algorithms}

\input{figures}

\input{intro}
\input{background}
\input{motivation}

\input{system}
\input{compilation}
\input{experiments}
\input{related}
\input{conclusion}
\end{sloppypar}

\pagebreak
\balance
\bibliographystyle{ACM-Reference-Format}
\bibliography{paper}

\end{document}

%% file: header.tex
\usepackage[noend]{algpseudocode}
\usepackage{algorithm}
\usepackage{bbm}
\usepackage{graphicx}
\usepackage{listings}
\usepackage[compress]{cleveref}
\usepackage{mathrsfs}
\usepackage{bm}
\usepackage{balance}
\usepackage{outlines}
\usepackage{svg}
\usepackage{grffile}
\usepackage[caption=false]{subfig}
\usepackage{xspace}
\usepackage{enumitem}
\usepackage{xcolor}
\usepackage{array}
\usepackage{nccmath}
\usepackage{multirow}
\usepackage{pifont}
\usepackage{soul}
\usepackage{caption}
\usepackage{upgreek}

\usepackage{xpatch}
\xpatchcmd{\NCC@ignorepar}{%
\abovedisplayskip\abovedisplayshortskip}
{%
\abovedisplayskip\abovedisplayshortskip%
\belowdisplayskip\belowdisplayshortskip}
{}{}

\graphicspath{./figures/}

\crefformat{section}{\S#2#1#3} %
\crefformat{subsection}{\S#2#1#3}
\crefformat{subsubsection}{\S#2#1#3}
\crefname{section}{Section}{Sections}
\Crefname{section}{Section}{Sections}
\crefname{figure}{Figure}{Figures}
\Crefname{figure}{Figure}{Figures}
\crefname{subfigure}{Figure}{Figures}
\Crefname{subfigure}{Figure}{Figures}
\crefname{algorithm}{Algorithm}{Algorithms}
\Crefname{algorithm}{Algorithm}{Algorithms}
\crefname{table}{Table}{Tables}
\Crefname{table}{Table}{Tables}
\crefrangelabelformat{subfigure}{#3#1#4--#5(\crefstripprefix{#1}{#2}#6}

\definecolor{emerald}{rgb}{0.31, 0.78, 0.47}

\newcommand{\revisioncolor}{black}
\newcommand{\revision}[1]{{\color{\revisioncolor} #1}}

\newcommand{\eat}[1]{}

\newcommand{\stitle}[1]{\vspace{0.1ex}\noindent{\bf #1}}

\newcommand{\preeq}{\vspace{0mm}\begin{small}}
\newcommand{\posteq}{\vspace{0mm}\end{small}}

\newcommand{\system}{\textsc{TQP}\xspace}

\usepackage{url}

\DeclareFixedFont{\ttb}{T1}{txtt}{bx}{n}{8}
\DeclareFixedFont{\ttm}{T1}{txtt}{m}{n}{8}

\usepackage{color}
\definecolor{deepblue}{rgb}{0,0,0.5}
\definecolor{deepred}{rgb}{0.6,0,0}
\definecolor{deepgreen}{rgb}{0,0.5,0}
\definecolor{purple}{rgb}{0.5,0,0.5}
\definecolor{gray}{rgb}{0.33,0.33,0.33}

\definecolor{dkgreen}{rgb}{0,0.6,0}
\definecolor{gray}{rgb}{0.5,0.5,0.5}
\definecolor{mauve}{rgb}{0.58,0,0.82}

\lstdefinelanguage{Python}{
	keywords={typeof, torch, nonzero, index_select, zeros_like, lt, masked_select, new, true, false, catch,def,val, function, return, null, catch, switch, var, shape,  while, do, else, case, break, override},
	keywordstyle=\color{blue}\bfseries,
	ndkeywords={class, export,extends, boolean, throw, implements, import, this, abstract, for, in, if},
	ndkeywordstyle=\color{dkgreen}\bfseries,
otherkeywords={+, =>,<=, ==, >,< , || , T},
	identifierstyle=\color{black},
	sensitive=false,
	comment=[l]{//},
	morecomment=[s]{/*}{*/},
	commentstyle=\color{purple}\ttfamily,
	stringstyle=\color{red}\ttfamily,
	morestring=[b]',
	morestring=[b]"
}

%% file: abstract.tex
\begin{abstract}

\begin{sloppypar}

The huge demand for computation in artificial intelligence (AI) 
is driving unparalleled investments in hardware and software systems for AI. 
This leads to an explosion in the number of specialized hardware devices, which are now offered by major cloud vendors. 
By hiding the low-level complexity through a tensor-based interface, tensor computation runtimes (TCRs) such as PyTorch %
allow data scientists to efficiently exploit the exciting capabilities offered by the new hardware. %
In this paper, we explore how database management systems can ride the wave of innovation happening in the AI space. 

\revision{We design, build, and evaluate Tensor Query Processor (\system): \system transforms SQL queries into tensor programs and executes them on TCRs. \system is able to run the full TPC-H benchmark by implementing novel algorithms for relational operators on the tensor routines.} 
At the same time, \system can support various hardware while only requiring a fraction of the usual development effort. 
Experiments show that \system can improve query execution time by up to \revision{10$\times$ %
over specialized CPU- and GPU-only systems.} %
Finally, \system can accelerate queries mixing ML predictions and SQL end-to-end, and deliver up to \revision{9$\times$} speedup over CPU baselines. 
 
 \end{sloppypar}
\end{abstract}

%% file: algorithms.tex
\newcommand{\GenericSortMergeJoinShort}{
    \begin{algorithm}[t!]
    \small
    \hspace{-10ex}\textbf{Input:} {\normalfont \emph{data}: input columns passed as an array of tensors.}
    
    \hspace{-10ex}\textbf{Output:} {\normalfont an array of tensors representing the join output.}
        \begin{algorithmic}[1]
            \State $\mathit{left}, \mathit{right} \gets \Call{getJoinKeyColumns}{\mathit{data}}$
            
            \hspace{-7ex}$\triangleright$ Sort join keys
            \State $\mathit{left}, \mathit{leftIdx} \gets \texttt{sort}(\mathit{left})$
            \State $\mathit{right}, \mathit{rightIdx} \gets \texttt{sort}(\mathit{right})$
            
            \hspace{-7ex}$\triangleright$ Build histograms for the left and right key columns
            \State $\mathit{leftHist}, \mathit{rightHist} \gets \texttt{bincount}(\mathit{left}), \texttt{bincount}(\mathit{right})$
            
            \hspace{-7ex}$\triangleright$  Compute the number of rows for each pair of matching keys
            \State $\mathit{histMul} \gets \texttt{mul}(\mathit{leftHist}, \mathit{rightHist})$

            \hspace{-7ex}$\triangleright$ Compute the prefix sums of histograms
            \State $\mathit{cumLeftHist} \gets \texttt{cumsum}(\mathit{leftHist}, \mathit{dim}=0)$
            \State $\mathit{cumRightHist} \gets \texttt{cumsum}(\mathit{rightHist}, \mathit{dim}=0)$
            \State $\mathit{cumHistMul} \gets \texttt{cumsum}(\mathit{histMul}, \mathit{dim}=0)$
            
            \hspace{-7ex}$\triangleright$ Initialize the output size and output offsets
            \State $\mathit{outSize} \gets \mathit{cumHistMul}[-1]$ 
            \State $\mathit{offset} \gets \texttt{arange}(\mathit{outSize})$
            
            \hspace{-7ex}$\triangleright$ Find the bucket of matching keys to which each output belongs
            \State $\mathit{outBucket} \gets \texttt{bucketize}(\mathit{offset}, \mathit{cumHistMul})$
            
            \hspace{-7ex}$\triangleright$ Compute the indexes from left and right in the join output
            \State $\mathit{offset}.\texttt{sub\_}(\mathit{cumHisMul}[\mathit{outBucket}] - \mathit{histMul}[\mathit{outBucket}])$
            
            \State $\mathit{leftOutIdx}\!\gets\!\mathit{leftIdx}[\mathit{cumLeftHist}[\mathit{outBucket}]\!-  \mathit{leftHist}[\mathit{outBucket}]\ \ \ \ \ \ \ \ \ \ $
            $\ \ +\  \texttt{div}(
                \mathit{offset},
                \mathit{rightHist}[\mathit{outBucket}],
                \mathit{rounding}=``\mathit{floor}")
            ]$
            \State $\mathit{rightOutIdx}\!\gets\! \mathit{rightIdx}[\mathit{cumRightHist}[\mathit{outBucket}]  -   $\emph{rightHist[outBucket}] \ \ \ \ \ \ \ \ \ \ \ \ \
            $\ \ +\  \texttt{remainder}(
                \mathit{offset},
                \mathit{rightHist}[\mathit{outBucket}])
            ]$
            
            \State \textbf{return} \Call{createOutput}{\emph{data}, \emph{leftOutIdx}, \emph{rightOutIdx}}
    \end{algorithmic}
    \caption{Sort-Based Join}
    \label{alg:GenericSortMergeJoinShort}
\end{algorithm}
}

\newcommand{\AggregationShort}{
    \begin{algorithm}[t!]
        \small
        \hspace{-10ex}\textbf{Input:} {\normalfont \emph{data}: input columns passed as an array of tensors.}
        
    \hspace{-13ex}\textbf{Output:} {\normalfont  the aggregation output as an array of tensors.}
        \begin{algorithmic}[1]
                \State $\mathit{grpByCols} \gets \Call{getGroupByColumns}{\mathit{data}}$
                
                \hspace{-7ex}$\triangleright$  Generate unique groups
                \State $\mathit{grps} \gets \texttt{cat}(\mathit{grpByCols}, \mathit{dim}=1)$
                
                \State $\mathit{grps}, \mathit{grpsInvIdx} \gets \texttt{sort}(\mathit{grps})$
                \State $\mathit{data} \gets [\mathit{col}[\mathit{grpsInvIdx}]\ \mathit{for}\ \mathit{col}\ \mathit{in}\ \mathit{data}]$
                \State $\mathit{grpsUnique},\!\mathit{invIdxs}\! \gets\!  \texttt{uniqueConsecutive}(\mathit{grps}, \mathit{inverse\!=\!True})$

                \hspace{-7ex}$\triangleright$ Evaluate the aggregation expression
                
                \State $\Return \text{ } [\Call{evaluate}{\mathit{data}, \mathit{grpsUnique}, \mathit{invIdxs}}]$

        \end{algorithmic}

        \caption{Aggregation}

        \label{alg:AggregationShort}
    \end{algorithm}
}

\newcommand{\GenericHashJoinShort}{
    \begin{algorithm}[t!]
        \small
        \hspace{-10ex}\textbf{Input:} {\normalfont \emph{data}: input columns passed as an array of tensors.}
        
    \hspace{-11ex}\textbf{Output:} {\normalfont an array of tensors representing the join output.}
        \begin{algorithmic}[1]

                \State $left, right \gets \Call{getJoinKeyColumns}{data}$
                
                \hspace{-7ex}$\triangleright$  Initialize the hash table size and indexes
                \State $mod \gets \Call{pow}{2, int(log(left.shape[0]))}$
                \State $leftIdx \gets \Call{arange}{left.shape[0]}$
                \State $rightIdx \gets \Call{arange}{right.shape[0]}$
                
                \hspace{-7ex}$\triangleright$  Compute the hash values for join keys
                \State $leftHash \gets \Call{hash}{left, mod}$
                \State $rightHash \gets \Call{hash}{right, mod}$
                
                \hspace{-7ex}$\triangleright$  Make the hash values positive
                \State $leftHash \gets \Call{fmod}{leftHash + mod, mod}$
                \State $rightHash \gets \Call{fmod}{rightHash + mod, mod}$
                
                \hspace{-7ex}$\triangleright$  Build histogram
                \State $hashBincount \gets \Call{bincount}{leftHash}$
                \State $maxNumRepeatingHash \gets \Call{max}{hashBincount}$
                
                \State $leftOutputIdx \gets \Call{empty}{0}$
                \State $rightOutputIdx \gets \Call{empty}{0}$
                
                \hspace{-7ex}$\triangleright$  Build and probe the hash table in an interleaved way
                \For {$i \in \Call{range}{maxNumRepeatingHash}$}
                    
                    \hspace{-3.5ex}$\triangleright$  Build the current hash table from left
                    
                    \If {$i == 0$}
                        \State $hashTable \gets \Call{full}{(mod + 1, ) -1}$
                    \Else
                        \State $hashTable.\Call{fill\_}{-1}$
                    \EndIf
                    
                    \State $hashTable.\Call{scatter\_}{0, leftHash, leftIdx}$
                    \State $leftIdxSct \gets \Call{maskedSelect}{hashTable, hashTable \ge 0}$
                    
                    \hspace{-3.5ex}$\triangleright$  Set mod as the hash value to abandon those already scattered indexes
                    \State $leftHash[leftIdxSct] \gets mod$
                    
                    \hspace{-3.5ex}$\triangleright$  Probe the current hash table from right
                    \State $leftCandIdx \gets hashTable[rightHash]$
                    \hspace{-3.5ex}$\triangleright$  Only non-negative values were scattered
                    \State $validKeyMask \gets leftCandIdx \ge 0$
                    
                    \hspace{-3.5ex}$\triangleright$  Get the scattered indexes and correspoding right indexes that have the same hash values
                    \State $validleftIdx \gets \Call{maskedSelect}{leftCandIdx, validKeyMask}$
                    \State $validrightIdx \gets \Call{maskedSelect}{rightIdx, validKeyMask}$
                    
                    \hspace{-3.5ex}$\triangleright$  Find the matching indexes and construct the output indexes
                    \State $matchMask \gets OnesLike(rightHash)$
                    \ForAll {$joinKeyIdx \in \Call{range}{self.numKeys}$}
                        \State $leftKeys \gets \Call{getLeftJoinKeyValues}{x, joinKeyIdx}$[validleftIdx]
                        \State $rightKeys \gets \Call{getRightJoinKeyValues}{x, joinKeyIdx}$[validrightIdx]
                        
                        \State $matchMask \gets \Call{logicalAnd}{matchMask, leftKeys == rightKeys}$
                        
                        \State $leftMatchIdx \gets \Call{maskedSelect}{validleftIdx, matchMask}$
                        \State $rightMatchIdx \gets \Call{maskedSelect}{validrightIdx, matchMask}$
                        \State $leftOutputIdx \gets \Call{cat}{(leftOutputIdx, leftMatchIdx)}$
                        \State $rightOutputIdx \gets \Call{cat}{(rightOutputIdx, rightMatchIdx)}$
                    \EndFor
                \EndFor
                
                \State $\Return \text{ } \Call{GenerateOutput}{x, outColumns, leftOutputIdx, leftOutputIdx}

        \end{algorithmic}

        \caption{Hash Join}

        \label{alg:GenericHashJoinShort}
    \end{algorithm}
}

%% file: figures.tex
\newcommand{\tpchvschip}{
    \begin{figure}  
    \vspace{-1ex}
    	\centering
    	\includegraphics[width=0.85\linewidth]{figures/tpc-vs-transistors.pdf}
    	\vspace{-3ex}
    	\caption{TPC-H scale factor 100 (non-cluster) performance compared to transistor count over the years.}
        \label{fig:tpc-vs-chips}
        \vspace{-5ex}
    \end{figure}
}

\newcommand{\dataRepFigure}{
    \begin{figure}  
    	\centering
    	\includegraphics[width=0.90\linewidth]{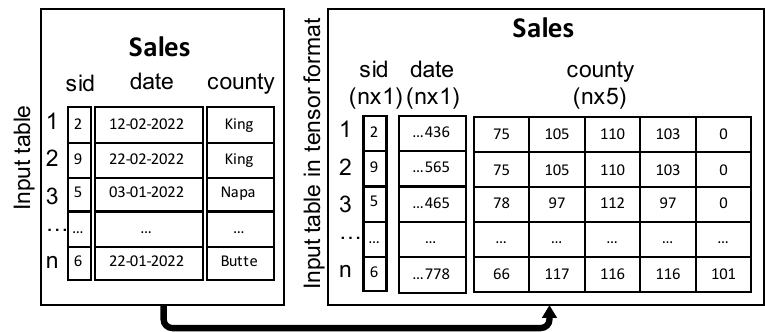}
    	\vspace{-2.25ex}
    	\caption{\system represents input tables in a columnar format with a 2d-tensor per column.}
    	\vspace{-3ex}
        \label{fig:data_rep}
    \end{figure}
}

\newcommand{\joinIllustration}{
    \begin{figure}[t!]
    \vspace{-0ex}
    	\centering
    	\includegraphics[width=0.90\linewidth]{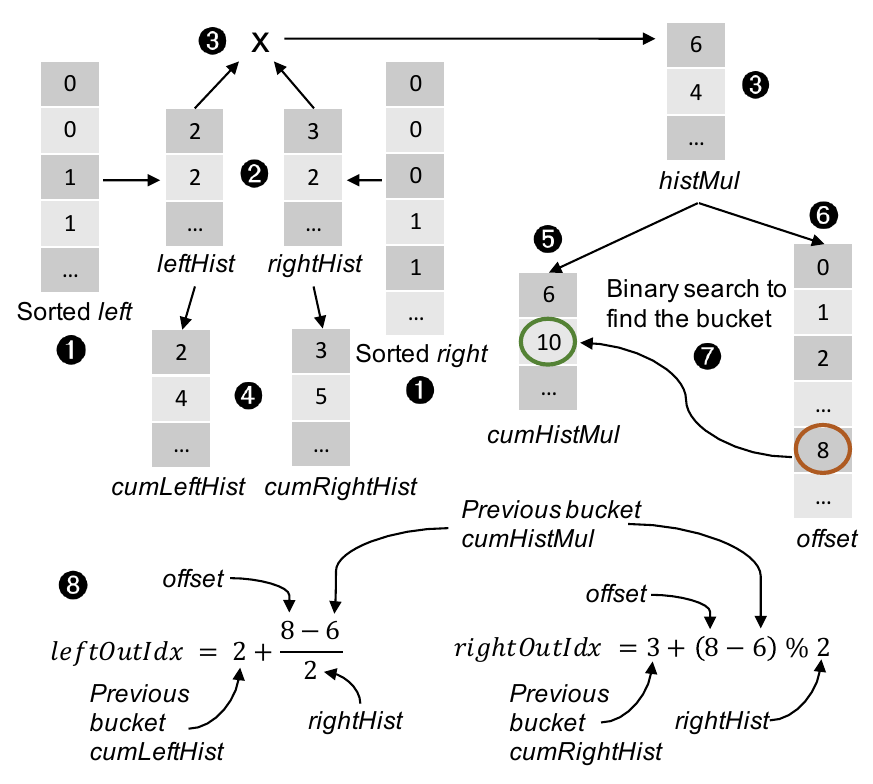}
    	\vspace{-3.25ex}
    	\caption{An example of the sort-based join implementation.}
        \label{fig:joinIllustration}
        \vspace{2ex}
    \end{figure}
}

\newcommand{\systemFigure}{
    \begin{figure}  
    \vspace{-0ex}
    	\centering
    	\includegraphics[width=\linewidth]{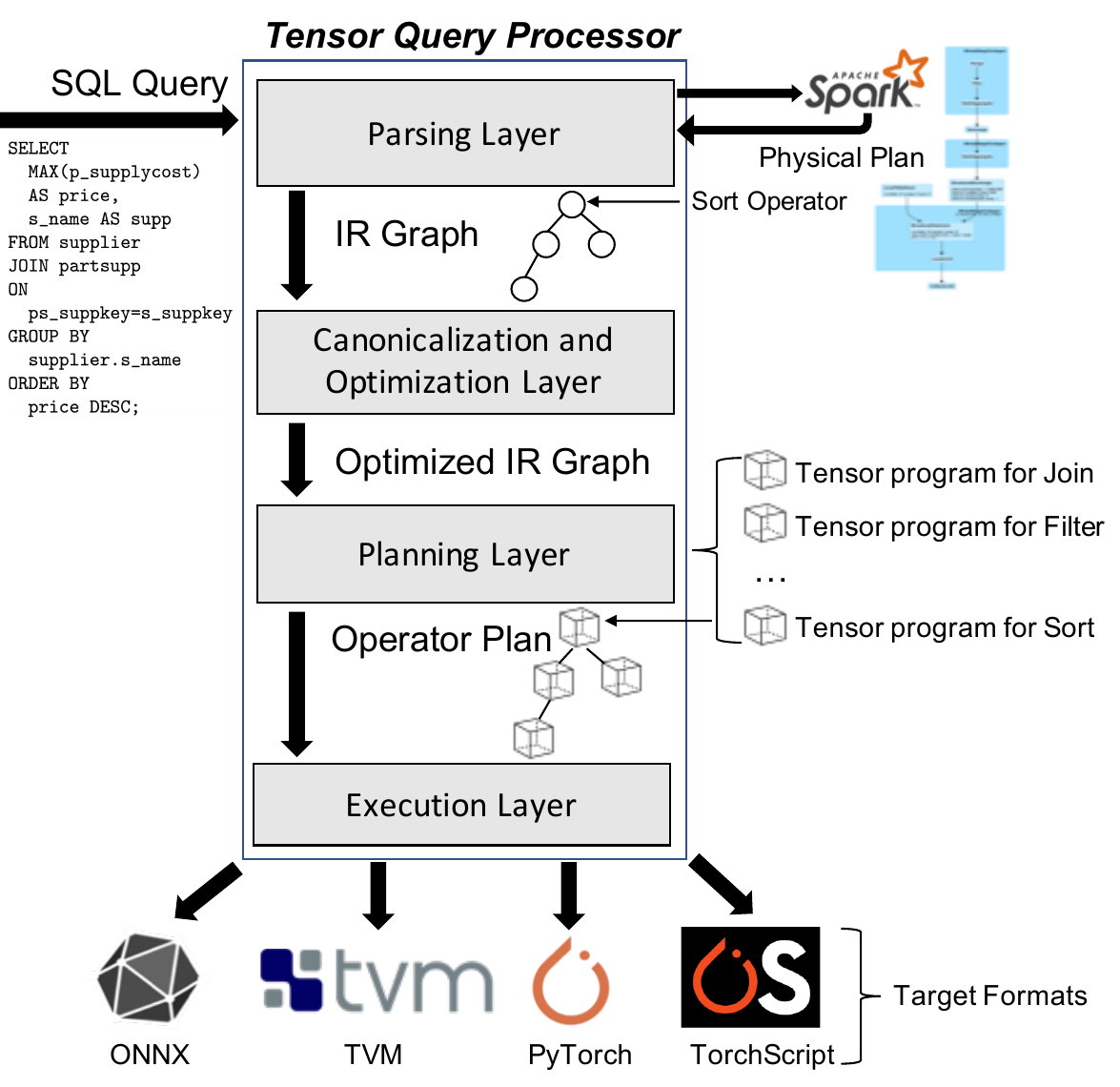}
    	\vspace{-4ex}
    	\caption{\system's compilation phase.}
        \label{fig:systemFigure}
        \vspace{-2ex}
    \end{figure}
}

\newcommand{\evalCpuGpuTableSmall}{
    \begin{table}[t!]
    \hspace{-2ex}
    \caption{\revision{Query execution time (in seconds) on the TPC-H benchmark (scale factor 1). Bold numbers highlight the best performance for the specific setup (CPU or GPU). We evaluate \system in two modalities: interpreted (\system) and compiled using TorchScript (TQPJ). N/A means the query execution did not finish because of an error. TQPJ currently does not support materialized views.}}
    \vspace{-2ex}
    \label{tbl:evalCpuGpuTable}
    \hspace{-1ex}
    \scalebox{0.97}{
    \small
    \begin{tabular}{c c c c c c c c c}
    \toprule
    \multirow{2}[2]{*}{Query} & \multicolumn{3}{c}{CPU (1 core)} & \multicolumn{5}{c}{GPU} \\
    \cmidrule(lr){2-5} \cmidrule(lr){6-9} & \multicolumn{1}{c}{Spark} & \multicolumn{1}{c}{DuckDB} & \multicolumn{1}{c}{\system} & \multicolumn{1}{c}{TQPJ} & \multicolumn{1}{c}{Blazing} & \multicolumn{1}{c}{Omnisci} & \multicolumn{1}{c}{\system} & \multicolumn{1}{c}{TQPJ}\\ 
    \midrule
    Q1 & 2.261 & \textbf{0.664} & \revision{7.535} & \revision{7.301} & 0.216 & 0.095 & \revision{0.027} & \revision{\textbf{0.026}}\\
    Q2 & 8.751 & \textbf{0.101} & \revision{0.629} & \revision{0.577} & 0.238 & 0.351 & \revision{0.039} & \revision{\textbf{0.028}}\\
    Q3 & 3.669 & \textbf{0.273} & \revision{1.154} & \revision{1.165} & 0.128 & 0.293 & \revision{0.027} & \revision{\textbf{0.024}}\\
    Q4 & 4.719 & \textbf{0.216} & \revision{1.050} & \revision{1.087} & 0.093 & 0.292 & \revision{0.020} & \revision{\textbf{0.018}}\\
    Q5 & 6.963 & \textbf{0.302} & \revision{2.459} & \revision{2.963} & 0.164 & 0.064 & \revision{0.048} & \revision{\textbf{0.042}}\\
    Q6 & 0.381 & 0.156 & \revision{0.143} & \revision{\textbf{0.073}} & 0.045 & 0.047 & \revision{0.003} & \revision{\textbf{0.002}}\\
    Q7 & 5.569 & \textbf{0.430} & \revision{2.236} & \revision{1.931} &  0.244 & 0.067 & \revision{0.042} & \revision{\textbf{0.035}}\\
    Q8 & 4.034 & \textbf{0.278} & \revision{2.460} & \revision{2.503} & 0.215 & 0.079 & \revision{0.050} & \revision{\textbf{0.039}}\\
    Q9 & 17.61 & \textbf{2.533} & \revision{4.518} & \revision{4.616} & 0.569 & \textbf{0.072} & \revision{0.105} & \revision{0.092}\\
    Q10 & 15.98 & \textbf{0.430} & \revision{1.168} &\revision{1.184} & 0.173 & 0.740 & \revision{0.057} & \revision{\textbf{0.052}}\\
    Q11 & 1.047 & \textbf{0.034} & \revision{0.476} & \revision{0.324} & N/A & 0.084 & \revision{0.016} & \revision{\textbf{0.009}}\\
    Q12 & 4.063 & \textbf{0.309} & \revision{0.976} & \revision{0.966} & 0.069 & 0.062 & \revision{0.025} & \revision{\textbf{0.021}}\\
    Q13 & 6.081 & \textbf{0.181} & \revision{9.379} & \revision{9.197} & 0.303 & \textbf{0.069} & \revision{0.153} & \revision{0.136}\\
    Q14 & 0.509 & 0.171 & \revision{0.124} & \revision{\textbf{0.096}} & 0.076 & N/A & \revision{0.007} & \revision{\textbf{0.005}}\\
    Q15 & 2.640 & 0.291 & \revision{\textbf{0.133}} & \revision{N/A} & N/A & \textbf{0.086} & \revision{ 0.129} & \revision{N/A} \\
    Q16 & 16.94 & \textbf{0.093} & \revision{3.664} & \revision{3.699} & N/A & 3.689 & \revision{ 0.320} & \revision{\textbf{0.301}}\\
    Q17 & 3.165 & \textbf{0.381} & \revision{2.303} & \revision{2.466} & 0.121 & 0.132 & \revision{ 0.061} & \revision{\textbf{0.051}}\\
    Q18 & 6.942 & \textbf{0.765} & \revision{2.245} & \revision{2.406} & 0.204 & 0.593 & \revision{0.053} & \revision{\textbf{0.048}}\\
    Q19 & 2.300 & \textbf{0.419} & \revision{1.577} &  \revision{1.316} & 0.188 & 0.058 & \revision{0.042} & \revision{\textbf{0.036}}\\
    Q20 & 4.232 & \textbf{0.276} & \revision{2.032} & \revision{1.975} & 0.149 & N/A & \revision{0.048} & \revision{\textbf{0.041}}\\
    Q21 & 12.39 & \textbf{0.932} & \revision{25.49} & \revision{24.25} & N/A & N/A & \revision{0.158} & \revision{\textbf{0.151}}\\
    Q22 & 3.919 & \textbf{0.069} & \revision{0.315} & \revision{0.296} & N/A & N/A & \revision{0.011} & \revision{\textbf{0.010}}\\
    \bottomrule
    \end{tabular}
    \normalsize
   }
    \vspace{-1ex}
    \end{table}
}

\newcommand{\evalScalabilityFigure}{
    \begin{figure*}[t!]
        \begin{center}
            \captionsetup{font={color=\revisioncolor}}
            \hspace{1.0em}
            \subfloat{\includegraphics[width=0.55\linewidth]{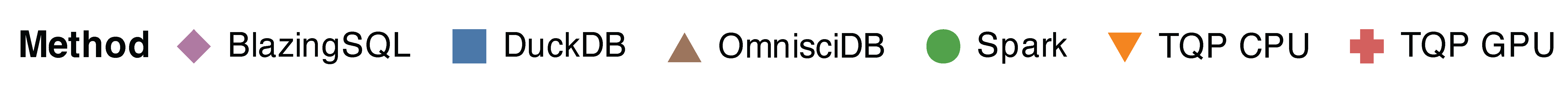}} \hfill%
            \vspace{-1.6em}
            \setcounter{subfigure}{0} 

            \subfloat[\revision{Query execution time over different numbers of cores.}]{\includegraphics[width=0.495\linewidth]{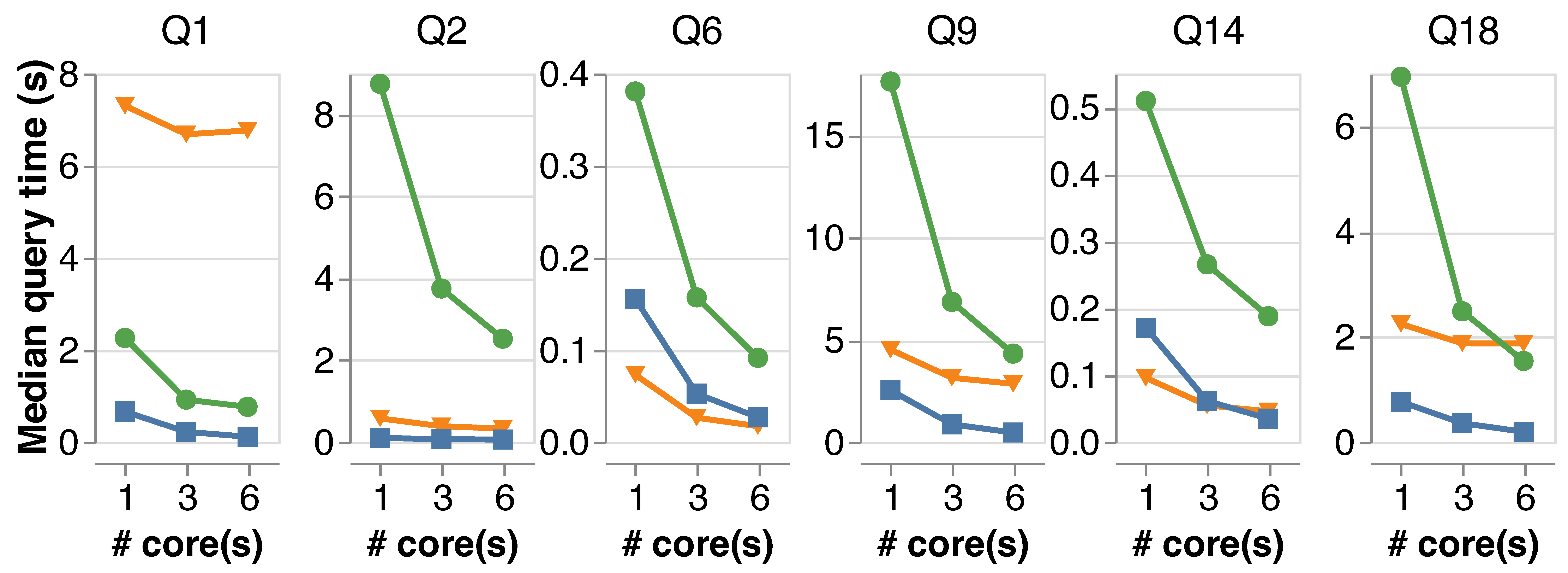}%
                \label{fig:evalScaleThreadFigure}}%
            \hfil
            \subfloat[\revision{Query execution time over different scale factors.}]{\includegraphics[width=0.505\linewidth]{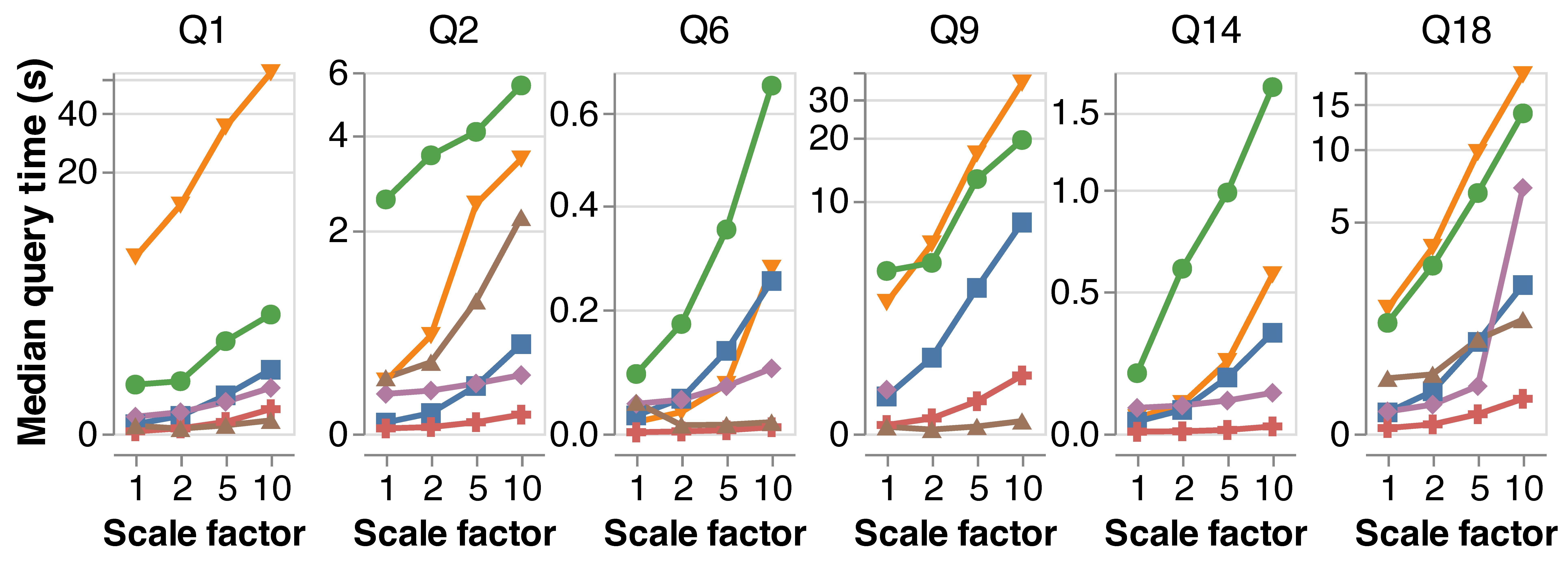}%
                \label{fig:evalScaleDatasetFigure}}%
        \end{center}
        \vspace{-2.6ex}
        \caption{\revision{Scalability on selected queries from TPC-H. For \system, we report the best time of the interpreted (PyTorch) and compiled (TorchScript) versions. In (a), the scale factor is 1. In (b), all CPU methods use 6 cores. BlazingSQL throws errors for Q9 at scale factors 2, 5, and 10. OmnisciDB does not support Q14. The y-axes in (b) are in (symmetric) log scale.}}
        \vspace{-2ex}
    \end{figure*}
}

\newcommand{\evalOpsBreakdownFigure}{
    \begin{figure*}[t!]
    \vspace{0ex}
        \begin{center}
            \captionsetup{font={color=\revisioncolor}}

            \subfloat[\vspace{-2ex}\revision{Query time breakdown for tensor operators on CPU}]{\includegraphics[width=0.51\linewidth]{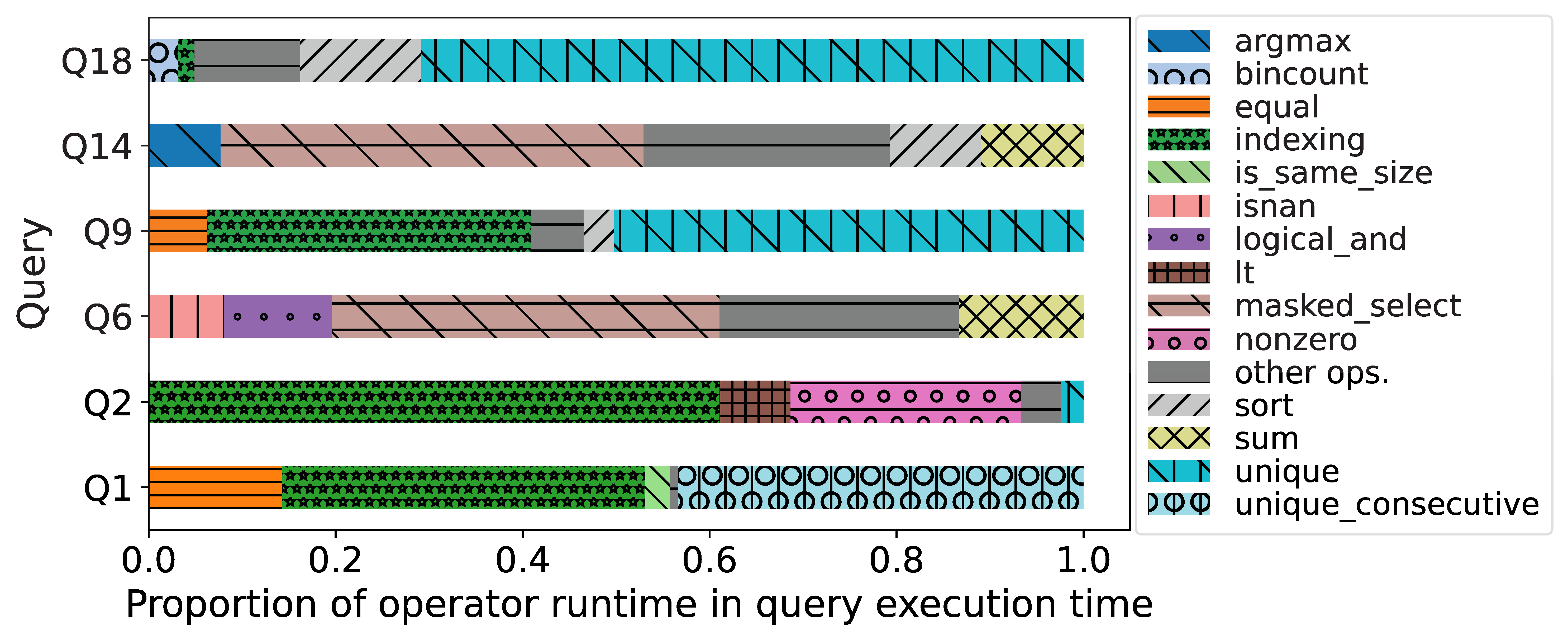}%
                \label{fig:evalOpsBreakdownCpuFigure}}%
            \hfil
            \subfloat[\revision{Query time breakdown for tensor operators on GPU}]{\includegraphics[width=0.488\linewidth]{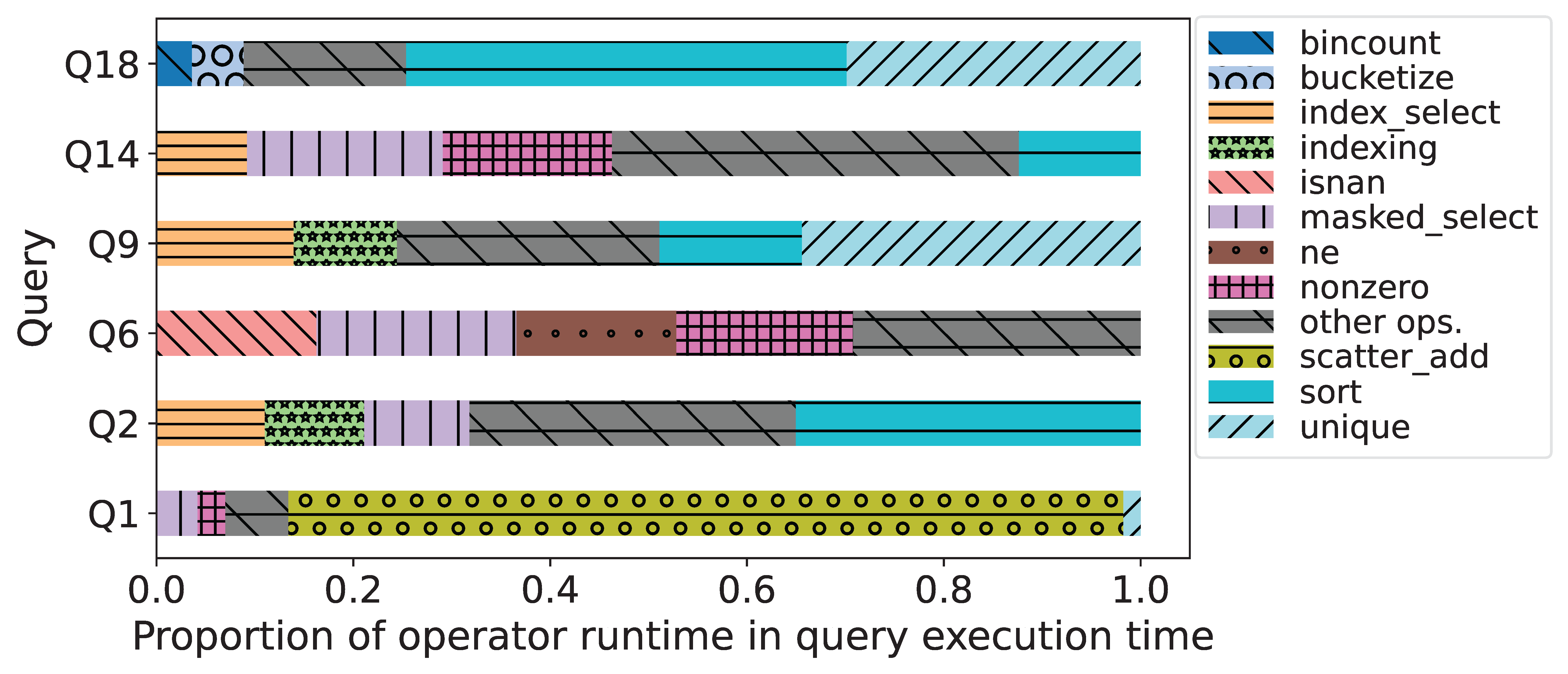}%
                \label{fig:evalOpsBreakdownCudaFigure}}%
        \end{center}
        \vspace{-1.2em}
        \caption{\revision{Query time breakdown for tensor operators for selected TPC-H queries at scale factor 10.}}
        \label{fig:evalOpsBreakdown}
        \vspace{-2.5ex}
    \end{figure*}
}

\newcommand{\evalOverheadFigure}{
    \begin{figure*}[t!]
    \vspace{-0ex}
        \begin{center}
            \captionsetup{font={color=\revisioncolor}}
            \hspace{1.1em}
            \subfloat{\includegraphics[width=0.575\linewidth]{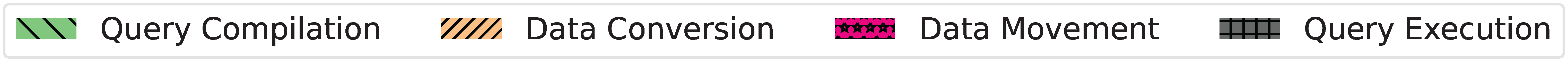}} \hfill%
            \vspace{-1.25em}
            \setcounter{subfigure}{0} 

            \subfloat[\revision{End-to-end execution breakdown on CPU}]{\includegraphics[width=0.485\linewidth]{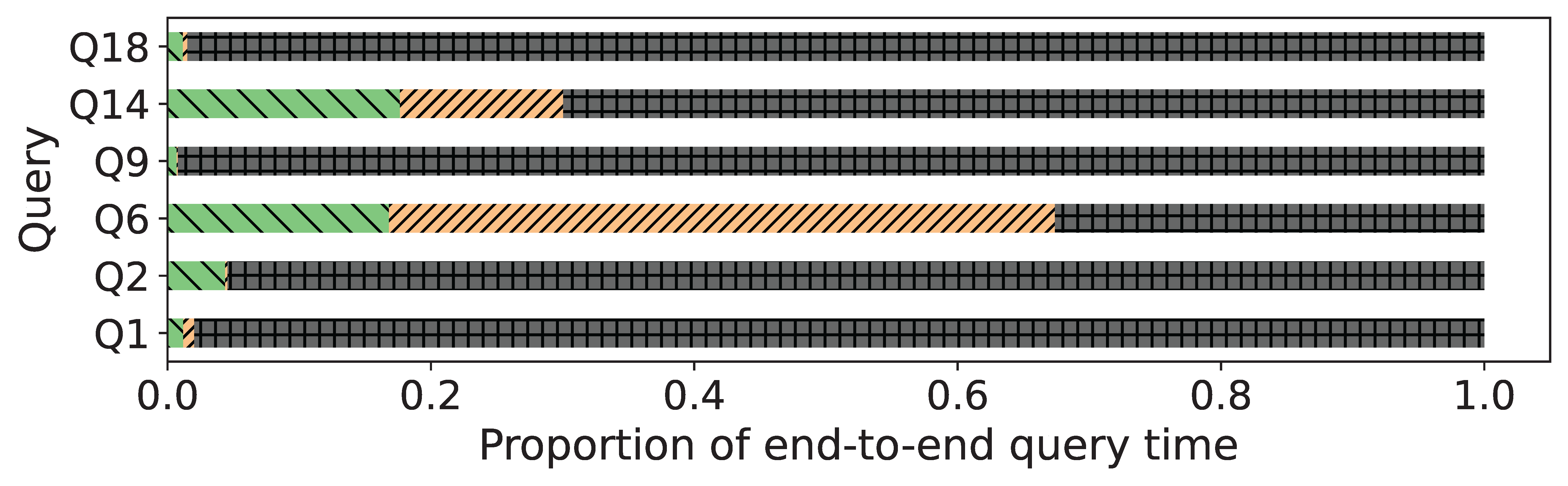}%
                \label{fig:evalOverheadFigureCpu}}%
            \hfil
            \subfloat[\revision{End-to-end execution breakdown on GPU}]{\includegraphics[width=0.485\linewidth]{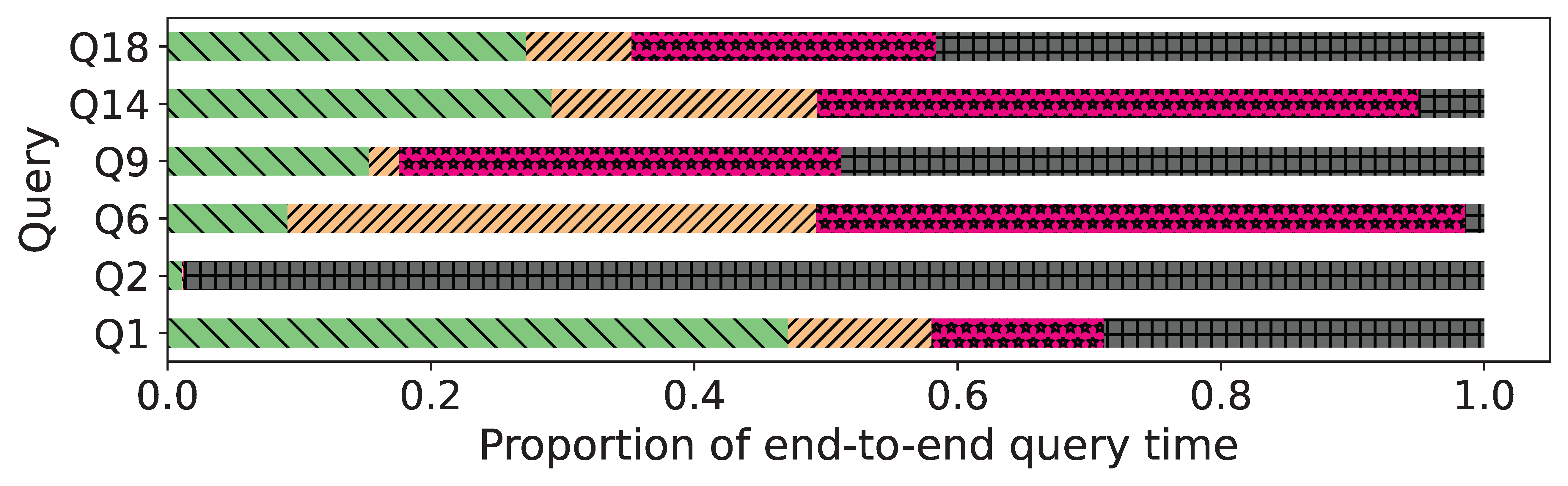}%
                \label{fig:evalOverheadFigureGpu}}%
        \end{center}
        \vspace{-1.3em}
        \caption{\revision{End-to-end breakdown (incl. all overheads, and w/o pipelining and caching) for selected queries at scale factor 10.}}
        \label{fig:evalOverhead}
        \vspace{-2ex}
    \end{figure*}
}

\newcommand{\evalOverheadTableSfTen}{
    \begin{table}[ht!]
    \hspace{-2ex}
    \caption{Overheads of \system for TPC-H (scale factor 10). Percentages represent the fractions over the corresponding end-to-end query time. Reported numbers are in seconds.} %
    \vspace{-2ex}
    \label{tbl:evalOverheadTable}
    \hspace{-1ex}
    \scalebox{0.97}{
    \small
    \begin{tabular}{c c c c c}
    \toprule
    \multirow{2}[2]{*}{TPC-H} & \multicolumn{2}{c}{CPU (1 core)} & \multicolumn{2}{c}{GPU} \\
    \cmidrule(lr){2-3} \cmidrule(lr){4-5} \multirow{1}[2]{*}{Query} & \multicolumn{1}{c}{Query} & \multicolumn{1}{c}{Data} & \multicolumn{1}{c}{Query} & \multicolumn{1}{c}{Data}\\
    & \multicolumn{1}{c}{Compilation} & \multicolumn{1}{c}{Conversion} & \multicolumn{1}{c}{Compilation} & \multicolumn{1}{c}{Movement}\\ 
    \midrule
    Q1 & 0.82 & 0.579 (0.85\%) & 2.506 & 0.695 (25\%)\\ 
Q2 & 0.607 & 0.026 (0.2\%) & 0.589 & 0.041 (0.08\%)\\ 
Q3 & 0.26 & 0.125 (1.5\%) & 0.254 & 0.279 (49\%)\\ 
Q4 & 0.174 & 0.237 (2.5\%) & 0.185 & 0.226 (38\%)\\ 
Q5 & 0.286 & 0.027 (0.17\%) & 0.244 & 0.269 (47\%)\\ 
Q6 & 0.142 & 0.427 (61\%) & 0.097 & 0.524 (54\%)\\ 
Q7 & 0.271 & 0.098 (0.63\%) & 0.294 & 0.309 (43\%)\\ 
Q8 & 0.303 & 0.083 (0.56\%) & 0.29 & 0.362 (49\%)\\ 
Q9 & 0.248 & 0.038 (0.11\%) & 0.255 & 0.561 (40\%)\\ 
Q10 & 0.203 & 0.071 (0.82\%) & 0.235 & 0.254 (34\%)\\ 
Q11 & 0.272 & 0.002 (0.043\%) & 0.267 & 0.031 (27\%)\\ 
Q12 & 0.172 & 0.378 (7.7\%) & 0.178 & 0.376 (43\%)\\ 
Q13 & 0.141 & 0.128 (0.57\%) & 0.139 & 0.601 (31\%)\\ 
Q14 & 0.145 & 0.101 (15\%) & 0.147 & 0.23 (64\%)\\ 
Q15 & 0.329 & 0.003 (0.46\%) & 0.378 & 0.001 (0.07\%)\\ 
Q16 & 0.232 & 0.116 (0.45\%) & 0.236 & 0.094 (1.9\%)\\ 
Q17 & 0.199 & 0.028 (0.11\%) & 0.211 & 0.182 (29\%)\\ 
Q18 & 0.231 & 0.068 (0.35\%) & 0.231 & 0.196 (32\%)\\ 
Q19 & 0.212 & 0.836 (21\%) & 0.226 & 1.226 (51\%)\\ 
Q20 & 0.252 & 0.16 (4.1\%) & 0.221 & 0.286 (52\%)\\ 
Q21 & 0.239 & 0.274 (0.14\%) & N/A & N/A\\ 
Q22 & 0.22 & 0.006 (0.2\%) & 3.009 & 0.02 (23\%)\\ 
    \bottomrule
    \end{tabular}
    \normalsize
   }
    \vspace{1ex}
    \end{table}
}

\newcommand{\evalHandCodedTable}{
    \begin{table*}[ht!]
    \hspace{-2ex}
    \caption{Query execution time (in seconds) on selected TPC-H queries (scale factor 10). \system Hand-Opt. uses hand-optimized tensor programs. %
    We use Torch, JIT, and TVM to refer to execution using PyTorch (interpreted), TorchScript (compiled), and TVM, respectively. Bold numbers highlight the best performance for the specific setup: CPU (1 core), CPU (6 cores), or GPU.}
    \vspace{-2.5ex}
    \label{tbl:evalHandCodedTable}
    \hspace{-1ex}
    \scalebox{0.94}{
    \small
    \begin{tabular}{c c c c c c c c c c c c c c c c c c c c}
    \toprule
    \multirow{3}[2]{*}{TPC-H Query} & \multicolumn{4}{c}{CPU (1 core)} & \multicolumn{4}{c}{CPU (6 cores)} &\multicolumn{4}{c}{GPU} \\
    \cmidrule(lr){2-5} \cmidrule(lr){6-9} \cmidrule(lr){10-13} & \multirow{2}[2]{*}{Best Baseline} & \multicolumn{3}{c}{\system Hand-Opt.} & \multirow{2}[2]{*}{Best Baseline} & \multicolumn{3}{c}{\system Hand-Opt.} & \multirow{2}[2]{*}{Best Baseline} & \multicolumn{3}{c}{\system Hand-Opt.} \\
    \cmidrule(lr){3-5} \cmidrule(lr){7-9} \cmidrule(lr){11-13} & & \multicolumn{1}{c}{Torch} & \multicolumn{1}{c}{JIT} & \multicolumn{1}{c}{TVM} & & \multicolumn{1}{c}{Torch} & \multicolumn{1}{c}{JIT} & \multicolumn{1}{c}{TVM} & & \multicolumn{1}{c}{Torch} & \multicolumn{1}{c}{JIT} & \multicolumn{1}{c}{TVM}\\
    \midrule
    Q1 & 6.54 (DuckDB) & \textbf{5.97}  & 6.89  & N/A & \textbf{1.1} (DuckDB) & 4.68  & 5.17  & N/A & 0.17 (OmnisciDB) & \textbf{0.13}  & 0.13  & N/A\\
    Q6 & 1.5 (DuckDB) & 0.87  & 1.18  & \textbf{0.24}  & 0.25 (DuckDB) & 0.66 & 0.71  & \textbf{0.12}  & 0.02 (OmnisciDB) & \textbf{0.01}  & 0.01 & 0.06\\
    Q9 & 45.11 (DuckDB) & 19.34  & \textbf{18.66}  & N/A & \textbf{7.75} (DuckDB) & 14.59  & 13.83  & N/A & \textbf{0.14} (OmnisciDB) & 0.45  & 0.44  & N/A\\
    Q14 & 1.7 (DuckDB) & 0.52  & 0.49  & \revision{\textbf{0.47}} & 0.33 (DuckDB) &  0.12  & \textbf{0.10}  & \revision{0.16} & 0.12 (BlazingSQL) & 0.01  & \textbf{0.01}  & \revision{0.30} \\
    \bottomrule
    \end{tabular}
    \normalsize
   }
    \vspace{-2.5ex}
    \end{table*}
}

\newcommand{\evalCostPerformanceFigure}{
    \begin{figure}
    	\centering
    	\vspace{0.5ex}
    	\includegraphics[width=1.0\linewidth]{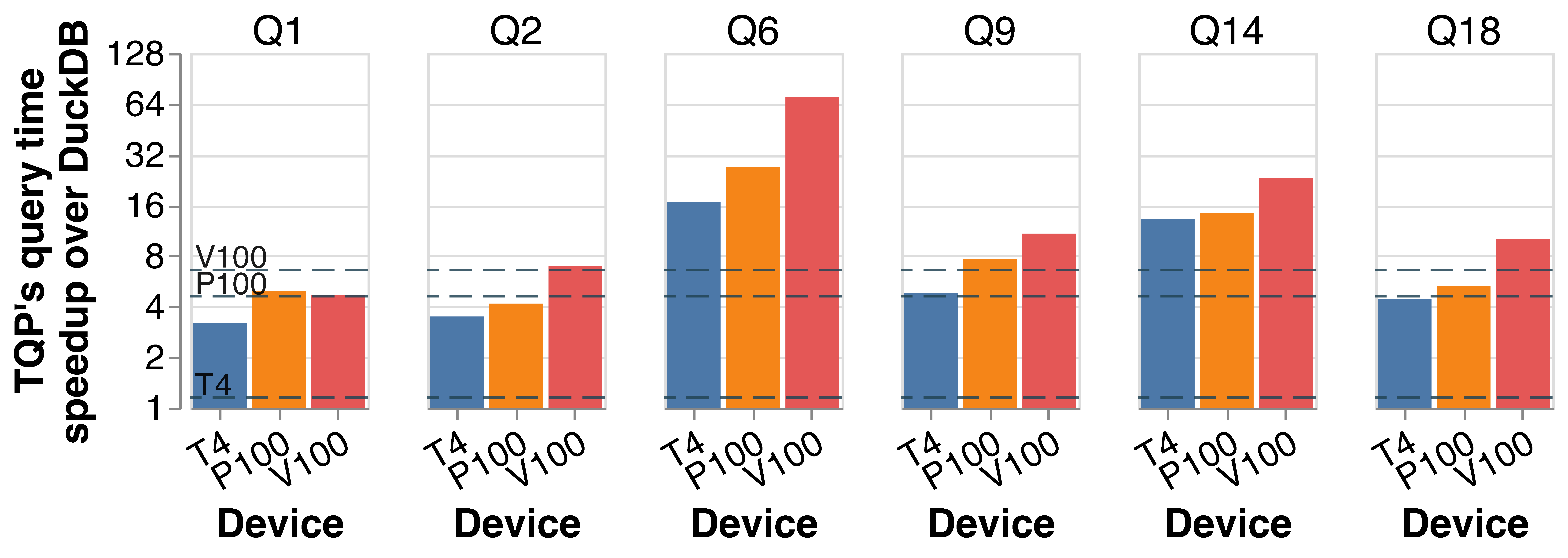}
    	\vspace{-5ex}
    	\caption{\revision{Cost/performance trade-off for \system on selected queries at scale factor 10.
    	We plot the speedups of \system on various GPUs (NVIDIA T4, P100 and V100) over DuckDB on a baseline CPU-only machine. The dashed lines represent the query time speedups required by the GPU executions to be more cost-effective compared to the DuckDB CPU baseline. %
    	}}
    	\vspace{-1ex}
        \label{fig:cost-perf}
    \end{figure}
}

\newcommand{\evalUtilizationFigure}{
    \begin{figure}  
        \vspace{-0.5ex}
    	\centering
    	\includegraphics[width=0.98\linewidth]{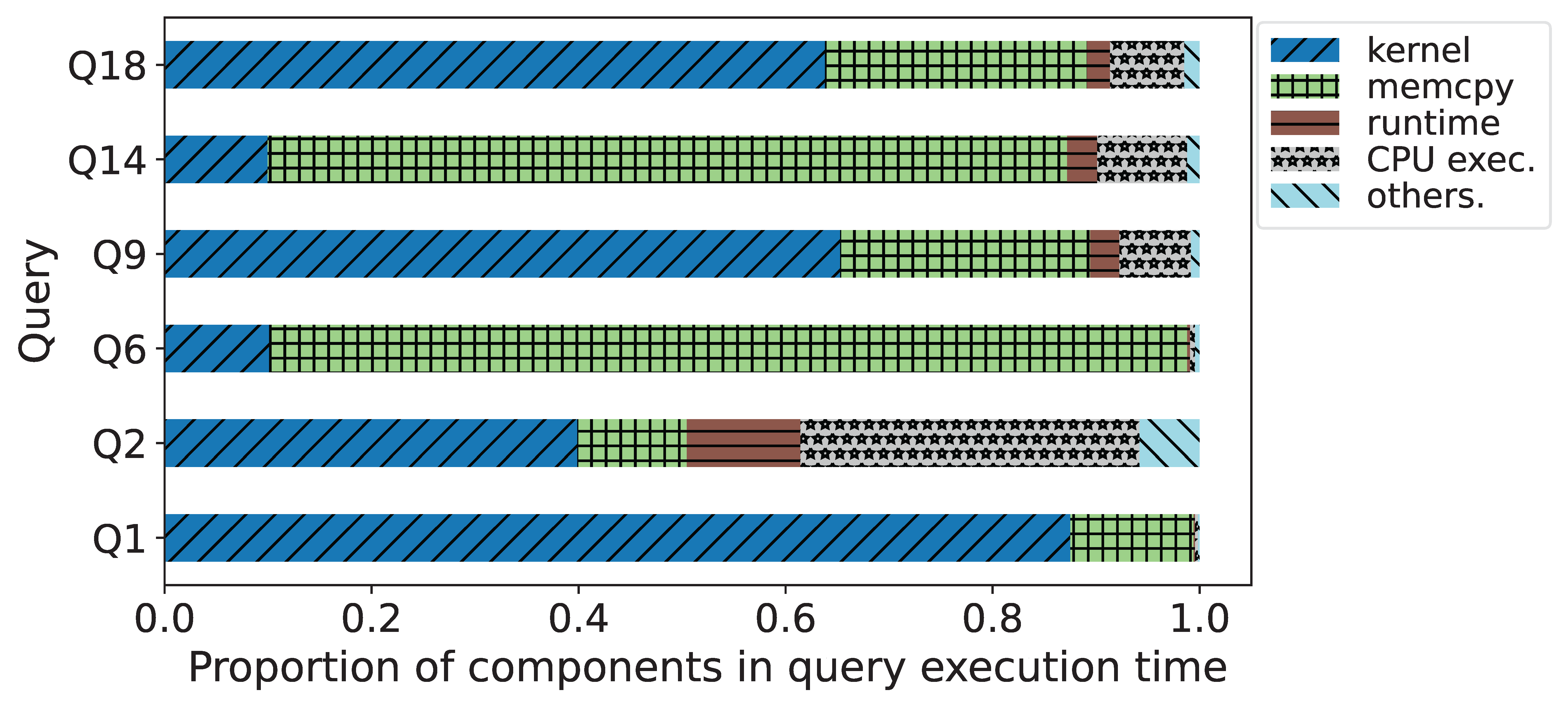}
    	\vspace{-2.5ex}
    	\caption{\revision{GPU utilization breakdown for selected TPC-H queries at scale factor 10. Utilization varies by query. Runtime is the time spent in scheduling the kernels.}}
        \label{fig:evalUtilizationFigure}
        \vspace{-1ex}
    \end{figure}
}

\newcommand{\evalPredictiveFigure}{
    \begin{figure}  
    	\centering
    	\hspace{-2ex}\includegraphics[width=0.825\linewidth]{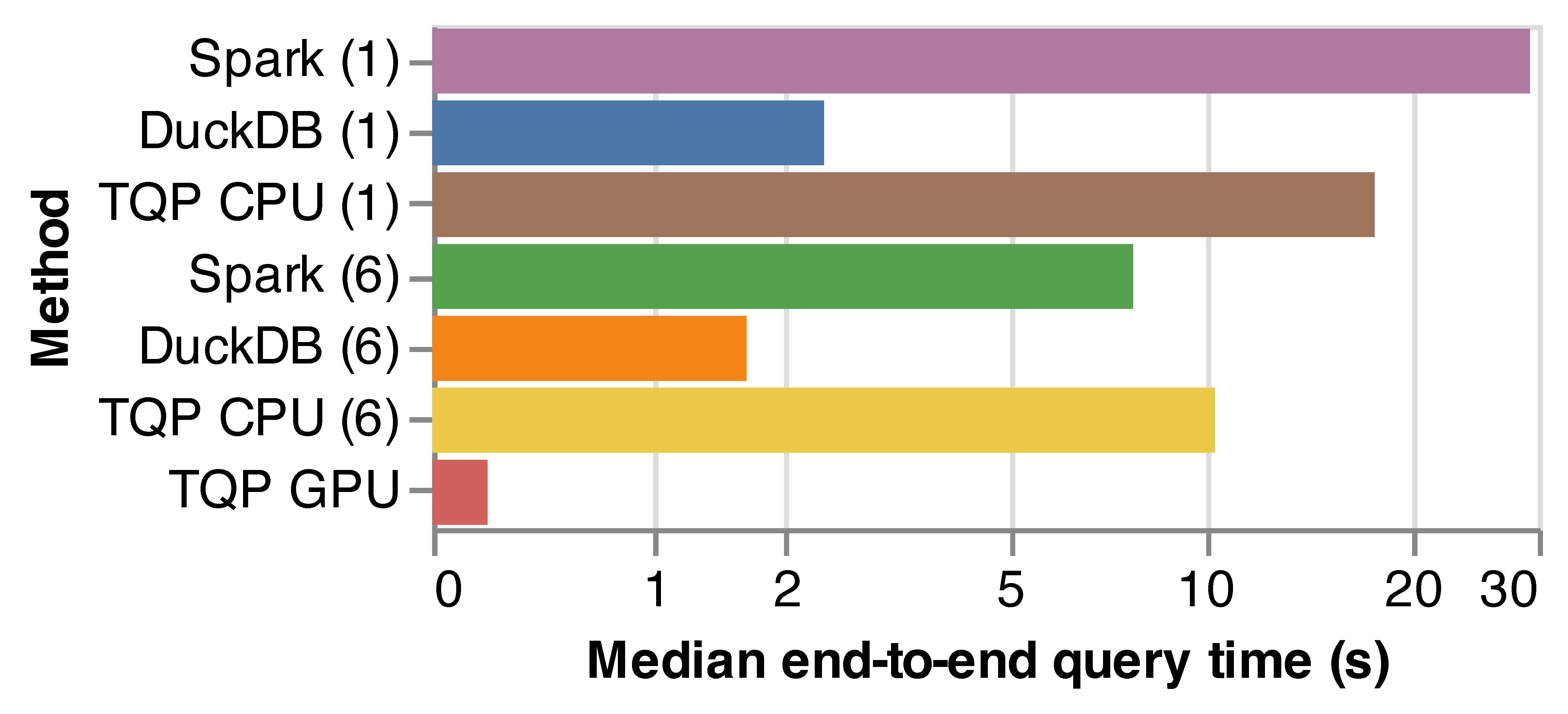}
    	\vspace{-3ex}
    	\caption{\revision{Query time on a query mixing ML prediction and relational operators. In parenthesis shows the number of CPU cores. The x-axis is in (symmetric) log scale.}}
        \label{fig:evalPredictiveFigure}
        \vspace{-1ex}
    \end{figure}
}

\newcommand{\evalPortabilityTable}{
    \begin{table*}[ht!]
    \vspace{-0ex}
    \hspace{-2ex}
    \caption{Query time (in milliseconds) of TPC-H Query 6 (scale factor 1) using the hand-optimized plan over different hardware and software backends. In parenthesis is the TCR used as well as the compilation stack (when applicable).}
    \vspace{-2.5ex}
    \label{tbl:evalPortabilityTable}
    \hspace{-1ex}
    \scalebox{0.96}{
    \small
    \begin{tabular}{c c c c c c}
    \toprule
    \multicolumn{1}{c}{Intel UHD Graphics 630} & \multicolumn{1}{c}{AMD Radeon Pro 5300M} &
    \multicolumn{1}{c}{NVIDIA K80} &
    \multicolumn{1}{c}{NVIDIA V100}
    &\multicolumn{1}{c}{TPU} & \multicolumn{1}{c}{Chrome}\\
    \multicolumn{1}{c}{(TVM on Metal)}& \multicolumn{1}{c}{(TVM on Metal)} &
    \multicolumn{1}{c}{(PyTorch)} &
    \multicolumn{1}{c}{(PyTorch)} &
    \multicolumn{1}{c}{(PyTorch on XLA)} & \multicolumn{1}{c}{(ORT on WASM)}\\ 
    \midrule
    62 & 17 & 5 & 1 & 25 & 1900\\
    \bottomrule
    \end{tabular}
    \normalsize
   }
    \vspace{-2ex}
    \end{table*}
}

\newcommand{\evalOperatorBreakdownTable}{
    \begin{table*}[ht!]
    \vspace{-5ex}
    \hspace{-2ex}
    \caption{\revision{Top-3 tensor operators that take the most time (incl. child operators) when executing selected queries from TPC-H (scale factor 1) on CPU and GPU. In parenthesis shows the fraction of the query execution time for the corresponding operator.}}
    \vspace{-2ex}
    \label{tbl:evalOperatorBreakdownTable}
    \hspace{-1ex}
    \scalebox{0.96}{
    \small
    \revision{
    \begin{tabular}{c c c c c c c}
    \toprule
    \multicolumn{1}{c}{TPC-H Query} & \multicolumn{3}{c}{\system CPU} & \multicolumn{3}{c}{\system GPU} \\
    \cmidrule(lr){1-1} \cmidrule(lr){2-4} \cmidrule(lr){5-7}
   Q1 & unique\_consecutive (0.43) & gen. indexing (0.39) & equal (0.14) & scatter\_add (0.84) & masked\_select (0.04) & nonzero (0.03) \\
Q2 & gen. indexing (0.61) & nonzero (0.25) & lt (0.08) & lt (0.32) & gen. indexing (0.32) & nonzero (0.32) \\
Q6 & masked\_select (0.41) & sum (0.13) & logical\_and (0.12) & masked\_select (0.20) & nonzero (0.18) & isnan (0.16) \\
Q9 & unique (0.47) & gen. indexing (0.35) & equal (0.06) & unique (0.29) & sort (0.19) & index\_select (0.12) \\
Q14 & masked\_select (0.45) & sum (0.11) & sort (0.10) & sort (0.22) & masked\_select (0.21) & nonzero (0.17) \\
Q18 & unique (0.70) & sort (0.13) & bincount (0.03) & sort (0.44) & unique (0.28) & bucketize (0.04) \\
    \bottomrule
    \end{tabular}
    \normalsize
   }}
    \vspace{-2ex}
    \end{table*}
}

\newcommand{\evalLOCTable}{
    \begin{table}[t!]
    \vspace{0.8ex}
    \caption{Lines of source code for implementing relational operators, excluding blank lines and comments.}
    \vspace{-2.5ex}
    \label{tbl:evalLOCTable}
    \scalebox{0.96}{
    \small
    \begin{tabular}{c c c c}
    \toprule
    \multirow{2}[1]{*}{System} & \multicolumn{3}{c}{Relational Operator} \\
    \cmidrule(lr){2-4} & \multicolumn{1}{c}{Hash Join} & \multicolumn{1}{c}{Sort-Based Join} & \multicolumn{1}{c}{Aggregation} \\
    \midrule
    \system (Various HW) & 148 & 182 & 104 (sort-based) \\
    Spark(CPU) & 706 & 1439 & 637 (sort-based) \\
    DuckDB (CPU) & 1415 & 877 & 1466 (hash-based) \\
    BlazingSQL (GPU) & 1628 & N/A & 1389 (hash-based) \\
    OmnisciDB (GPU) & 10141 & N/A & 2416 (hash-based) \\
    \bottomrule
    \end{tabular}
    \normalsize
   }
    \end{table}
}

%% file: intro.tex
\vspace{-0.5ex}
\section{Introduction}
\label{sec:introduction}

DBMS vendors have delivered constant performance improvement for decades by evolving software to keep up with Moore's law while influencing hardware development through close relationships with manufacturers. 
While data volumes and demand for analytics are growing faster than ever \cite{data-volume-growth}, the performance improvement on CPU has slowed down \cite{moore-end}. 
However, the count of processor transistors has continued to grow over the last decade, as hardware manufacturers adopted first multi-core CPU architectures and then augmented their computing platforms with specialized components such as GPUs, FPGAs, compression and encryption chips, DSPs, and neural-network (NN) accelerators. %
Although DBMS builders have taken advantage of multi-core and SIMD instructions effectively \cite{database-simd,database-simd-revisited,multicore-join}, the explosion in the number of specialized hardware components, each with different characteristics and programming abstractions, makes it challenging to support all the exciting capabilities that these new powerful devices can offer.

On the other hand, the huge demand for computation in artificial intelligence (AI) \cite{ai-memory-wall}, combined with the market fever for AI, is driving unparalleled investments in new hardware and software for AI. 
Hardware makers (e.g., Intel~\cite{habana}, Apple~\cite{neural-engine}, Xilinx~\cite{xilinx-ai}, AMD~\cite{rocm}), cloud vendors (e.g., Amazon~\cite{inferentia}, Microsoft~\cite{brainwave}, Google~\cite{tpu}), startups (e.g., Graphcore~\cite{graphcore}, Sambanova~\cite{sambanova},  Cerebras~\cite{cerebras}), and car companies like Tesla~\cite{d1} are investing heavily in this space. 
Venture capitals alone are pouring nearly \$2B a quarter on special hardware for AI, aiming for a market expected to exceed \$200B a year by 2025 \cite{ai-hw-market}. 
On the software side, companies and open source communities are rallying behind a small number of big efforts (e.g., PyTorch~\cite{pytorch-ecosystem}, TensorFlow~\cite{tensorflow2}, TVM~\cite{tvm}). 
The combination of investments in specialized hardware and large software communities focusing on performance allows these efforts to thrive. 
\revision{Our realization is that
the ML community has made hardware accelerators accessible to nonspecialists (e.g., data scientists). 
The fact that the most popular ML frameworks are open-source, creates a virtuous cycle whereby any hardware vendor interested in the ML space must support these frameworks well to get adoption. 
At the same time, these large open source communities successfully tackle the labor-intensive problem of providing specialized kernels for various hardware, e.g., a month after Apple M1 was announced, TVM outperformed Apple's CoreML by 2$\times$~\cite{tvm-on-m1}. 
Hardware vendors can directly improve the kernels' performance or the hardware itself~\cite{pytorch-intel,pytorch-metal,pytorch-amd}. 
This further helps adoption since the performance improves at each new software and hardware release.}

We argue that the best path forward for analytical DBMSs is to embrace this tectonic shift and take advantage of the groundswell of new hardware and software targeting AI workloads. 
To demonstrate the viability of this idea, we propose and prototype a new query processor which runs SQL queries atop tensor computation runtimes (TCRs) such as PyTorch, TVM, and ONNX Runtime~\cite{onnx-runtime}.
We name our prototype \emph{Tensor Query Processor} (\system). %
\revision{\system transforms a SQL query into a tensor program and executes it on TCRs.} 
To our knowledge, \system is the first query processor built atop TCRs. 
Careful architectural and algorithmic design enables \system to: 
(1) deliver significant \emph{performance} improvements over popular CPU-based data systems, and match or outperform custom-built solutions for GPUs; 
(2) demonstrate \emph{portability} across a wide range of target hardware and software platforms; and 
(3) achieve all the above with \emph{parsimonious} and sustainable \emph{engineering effort}. 

The above might appear surprising as specialized hardware accelerators are notoriously hard to program, requiring much customization to extract the best performance. 
Furthermore, their programming abstractions differ sufficiently to make our goals of \revision{\emph{performance} (G1), \emph{portability} (G2), and \emph{parsimonious engineering effort} (G3)} seemingly hard to reconcile. 
However, the key is a compilation layer and a set of novel algorithms, which can map the classical database abstraction to the prevalent one in machine learning (ML), i.e., \emph{mapping relational algebra to tensor computations}. 
This allows us to free-ride on existing labor-intensive efforts from the ML community to port and optimize TCRs across all the new specialized hardware platforms. 
The initial performance results are encouraging. 
On GPU, \system is able to \revision{outperform open-source GPU databases} in terms of query execution time. 
On CPU, \system outperforms Spark~\cite{spark}, and it is comparable to a state-of-the-art vectorized engine, DuckDB~\cite{duckdb}, for several queries. 
Furthermore, when ML and SQL queries are used in concert, \system is able to provide end-to-end acceleration for a \revision{9$\times$} speedup over CPU baselines.

Pursuing our goals of \emph{portability} and \emph{parsimonious engineering effort}, we make a deliberate decision to target existing tensor APIs rather than customize lower-level operators.
This decision potentially leaves some performance on the table but leads to a very sustainable long-term play, as \system benefits from any performance enhancement and optimization added to the underlying \revision{software and hardware~(e.g., \cite{pytorch-intel})}. 
To validate this proposition, we run \system on several different hardware settings: from CPUs, to discrete GPUs, to integrated GPUs (Intel and AMD), to NN-accelerators (TPUs~\cite{tpu}), and web browsers.
Furthermore, TQP is able to run the full TPC-H benchmark on both CPU and GPU with just \revision{around 8,000} lines of code---this is quite an achievement considering that until 2021 no GPU database was able to run all the 22 TPC-H queries~\cite{rateupdb}.

\stitle{Contributions.}
This paper makes the following core contributions:
\begin{itemize}[itemsep=1pt, topsep=0.5pt, leftmargin=12pt]
    \item We propose Tensor Query Processor (\system) that comprises a collection of algorithms and a compiler stack for transforming relational operators into tensor computations.
    \item With \system, we demonstrate that the tensor interface of TCRs is expressive enough to support all common relational operators.
    \item We evaluate the \system approach extensively against state-of-the-art baselines on the TPC-H benchmark.
\end{itemize}

\stitle{Organization.}
\cref{sec:background} introduces some background on TCRs. 
\cref{sec:motivation} summarizes the challenges and the design choices we make.
\cref{sec:surakav} introduces \system, and \cref{sec:algorithms} describes the algorithms used to implement several key relational operators with tensor programs. 
Experiments are in \cref{sec:experiments}.
 Related works are in \cref{sec:related}.
The paper is concluded by \cref{sec:conclusion}.

%% file: background.tex
\vspace{-1ex}
\section{Background}
\label{sec:background}

In this section, we summarize the system support for tensor computation (\cref{sec:tcr}), and provide a taxonomy of the tensor operations used throughout the paper (\cref{sec:tensor-operations}).

\vspace{-1.5ex}
\subsection{Tensor Computation Runtimes (TCRs)}
\label{sec:tcr}

The last years have witnessed an increase in the popularity of ML models based on NNs~\cite{dl-book}.
While in the heydays, these models were implemented manually in C++, data scientists \revision{now} can take advantage of several open-source ML frameworks simplifying the authoring and deployment of NN models.
TensorFlow~\cite{tensorflow} and
PyTorch~\cite{pytorch} are considered the most popular of such frameworks.

ML frameworks follow a common architecture: at the top, they have a \emph{high-level} Python API\footnote{Note that TCRs allow implementation in other languages too (e.g., Java~\cite{torch-java}, Rust~\cite{torch-rust}, C\#~\cite{torch-sharp}). Python is however the default language of choice by data scientists.} where data is commonly represented as multi-dimensional arrays called \emph{tensors}, while computation is expressed as a composition of \emph{tensor operations} embedded into the Python language.
At the lower level, they have a \emph{runtime} and a \emph{dispatcher}/\emph{compiler} allowing to run the tensor operations over different hardware backends such as CPU, GPU, custom ASICs, and using single node execution, distributed~\cite{distributed-pytorch}, or mobile/edge~\cite{google-tensor}.

Modern ML frameworks allow running computation in an \emph{interpreted mode} (often referred to as \emph{eager execution}), or in a \emph{compiled mode} (\emph{graph execution}), enabling code optimizations such as common sub-expression elimination, operator fusion, code generation~\cite{eager-vs-graph}, and removing Python dependency~\cite{tvm-passes,tvm-codegen}.
Interpreted vs. compiled execution is a popular dichotomy in query processing system implementations~\cite{vectorized-vs-compiled}. 
ML frameworks allow both modalities and we explore the trade-offs involved when using one vs. another, and the current limits of tensor compilers in \cref{sec:experiments}.

We will refer to ML frameworks, runtimes~\cite{onnx-runtime,tensor-rt}, and compilers as tensor computation runtimes (TCRs) in the rest of the paper.

\vspace{-1ex}
\subsection{Tensor Operations}
\label{sec:tensor-operations}

TCRs provide hundreds of tensor operations. 
We provide a brief summary of the operators used in \system, organized by category\footnote{Since \system is currently built on top of PyTorch, from now on we will use the PyTorch naming convention. Note that similar tensor operations can be found on other TCRs. Additionally, here we take the freedom to provide a different taxonomy than the one found in the PyTorch documentation~\cite{pytorch-tensor-doc} and in our previous work~\cite{hummingbird-vision}.}.

\begin{description}[style=multiline,wide,nosep]
    \item[Creation.] This category contains all operations used to create tensors, e.g., \texttt{from\_numpy}, fill a tensor with specific elements (\texttt{zeros}, \texttt{ones}, \texttt{empty}, \texttt{fill}, \texttt{arange}) or create a tensor using the same shape of another tensor (\texttt{zeros\_like}, \texttt{ones\_like}).
    \item[Indexing and slicing.]
    This category involves operations for selecting one or more elements of a
    tensor using the square bracket notation, or using indexing (\texttt{index\_select}), a mask (\texttt{masked\_select}), or a range (\texttt{narrow}). 
    \item[Reorganization.] This category includes  \texttt{reshape}, \texttt{view}, and \texttt{squeeze} that reorganize the shape of a tensor (eventually by changing only its metadata). \texttt{gather}, \texttt{scatter} reorganize the elements of a tensor using an index, while \texttt{sort} sorts its elements.
    \item[Comparison.] 
    \texttt{eq}, \texttt{lt}, \texttt{gt}, \texttt{le}, \texttt{ge}, \texttt{isnan} are operators in this category. Other operations are
    \texttt{where} that implements conditional statements, and \texttt{bucketize} that implements binary search.
    \item[Arithmetic.] \texttt{add}, \texttt{mul},
    \texttt{div}, \texttt{sub}, \texttt{fmod}, \texttt{remainder} are in this category. We also include logical operators such as \texttt{logical\_and}, \texttt{logical\_or}, \texttt{negative}, and shift operations.
    \item[Join.] This category allows to \texttt{concat} or \texttt{stack} multiple tensors.
    \item[Reduction.]
    This category contains operations for calculating simple aggregates (\texttt{sum}, \texttt{max}, \texttt{min}, \texttt{mean}), aggregates over groups (\texttt{scatter\_add}, \texttt{scatter\_min}, \texttt{scatter\_max}, \texttt{scatter\_mean}), logical reductions (\texttt{all}, \texttt{any}), as well as operations to build histograms
    (\texttt{bincount}, \texttt{histc}), \texttt{nonzero} (returning the indexes of non-zero elements), \texttt{unique} and \texttt{unique\_consecutive}.

\end{description}

%% file: motivation.tex
\section{Query Processing on TCRs}
\label{sec:motivation}

In this section, we summarize %
the  
challenges (\cref{sec:challenges}) and the design principles we commit to (\cref{sec:design}) when building \system. 
First, we show how relational operators can be implemented using tensor programs with an example (\cref{sec:tensor-program-example}).

\vspace{-1ex}
\subsection{Relational Operators as Tensor Programs}
\label{sec:tensor-program-example}

TCRs operate over data represented as tensors. Tensors are arrays of arbitrary dimensions containing elements of the same data type. 0d-tensors are referred to as \emph{scalars},  
1d-tensors as \emph{vectors}, and 2d-tensors as \emph{matrices}. 
For a tensor of $n$ dimensions, its \emph{shape} is a $n$-tuple where each element $i \in \{0, 1, \dots, n\}$ specifies the size of the $i$-dimension. For example, a matrix with 10 rows and 5 columns is a 2d-tensor of shape (10, 5). 
This paper only considers dense tensors where each element is explicitly stored in memory. 
\eat{TCRs often support \emph{sparse} tensors as well, where only non-zero elements are explicitly stored.
From our experience support for sparse tensor operations is however still limited. }%

ML practitioners implement programs (NNs) as a composition of operations over tensors.
While relational operations are commonly expressed as queries in a standalone language (e.g., SQL), tensor operations are embedded in a host language (e.g., Python), which is used to implement control flows and etc. 
Next, we introduce examples of implementing a filter using tensors. 

Let us assume that we want to implement a simple filter condition over the {\sc l\_quantity} column of the {\sc lineitem} table: {\sc where l\_quantity < 24}. 
First, we can represent {\sc l\_quantity} as a 1d-tensor of floating point numbers.
We can then use the \texttt{lt} (\texttt{l}ess \texttt{t}han) operator to implement the filter condition (line 1 of Listing~1). 
\texttt{lt} generates a boolean mask which is then used as a parameter of the \texttt{masked\_select} operator to generate the filtered version of the {\sc l\_quantity} column vector (line 2 of Listing~1). 
\label{lst:filter_bm}
\vspace{-2ex}
\begin{lstlisting}[caption=Filter implementation using bitmaps.]
mask = torch.lt(l_quantity, 24)
output = torch.masked_select(l_quantity, mask)
\end{lstlisting}
\vspace{-1ex}
\noindent This implementation is almost identical to the Bitmap-based representation~\cite{filter-representation} of filters in vectorized query processors~\cite{vix,blu}. 
On CPU, TCRs have SIMD implementations for several condition and intersection operators. 
An alternative is to use indexes rather than masks. This is commonly referred to as Selection Vector representation~\cite{filter-representation,vectorwise},  \revision{and can be similarly implemented using tensor operators \texttt{lt}, \texttt{nonzero}, and \texttt{index\_select}}.

Listing 2 shows another implementation. %
Here, we iterate over all the elements of the input tensor and use a Python conditional statement. This implementation does not take advantage of any tensor operation beyond creating the output tensor. %

\vspace{-2ex}
\begin{lstlisting}[caption=Filter implementation using Python control flow.]
output = torch.zeros_like(l_quantity), j = 0
for i in range(l_quantity.shape[0]):
    datum = l_quantity[i]
    if datum < 24:
        output[j] = datum, j = j + 1
output = output[:j, :]
\end{lstlisting}
\label{lst:filter_for}

\begin{table}[!b]
\vspace{-2ex}
\caption{Execution times of filter over $\sim$6M elements in interpreted (Torch) and compiled (TorchScript) modes.}
\vspace{-2.75ex}
\label{tbl:filter_example}
\scalebox{0.90}{
\small
\begin{tabular}{cccccccccc}
\toprule
\multirow{2}[1]{*}{Implementation}
 &
\multicolumn{2}{c}{CPU} & \multicolumn{2}{c}{GPU}\\
\cmidrule(lr){2-3} \cmidrule(lr){4-5}
 & \multicolumn{1}{c}{Torch} & \multicolumn{1}{c}{TorchScript} & \multicolumn{1}{c}{Torch} & \multicolumn{1}{c}{TorchScript} \\ 
\midrule
Bitmap & 
 \multicolumn{1}{c}{36.6ms} & \multicolumn{1}{c}{36.6ms} & \multicolumn{1}{c}{2.9ms} & \multicolumn{1}{c}{2.9ms} \\
 Python & 
 \multicolumn{1}{c}{23s} & \multicolumn{1}{c}{22.7s} & \multicolumn{1}{c}{200.3s} & \multicolumn{1}{c}{200s} \\
\bottomrule
\end{tabular}
}
\end{table}

\vspace{-1ex}
Table~\ref{tbl:filter_example} shows the performance of the two implementations. %
The implementation using Python control flow is considerably slower. 
, and GPU execution of Python control flow is slower than CPU execution. 
This result highlights one of the design choices (\cref{sec:design}) we make in \system: avoid the use of data-dependent code in Python.

\vspace{-2ex}
\subsection{Challenges}
\label{sec:challenges}

Implementing a query processor on TCRs requires overcoming several challenges. 
After all, TCRs are built for authoring and executing NN models, not relational queries.

\begin{description}[style=multiline,wide,nosep]
    \item[C1:\!\!]\emph{Expressivity}. Relational queries can contain filters with fairly complex expressions (e.g., {\sc like, in}), sub-queries, group-by aggregates, joins (e.g., natural, anti, semi, outer), etc. It is not clear whether the tensor operations currently available in TCRs are enough to support all these relational operators. %
    \item[C2:\!\!]\emph{Performance}. Even if a relational operator is implementable using tensors, this does not automatically lead to good performance, as the example in Listing~2 suggests. %
    \revision{In fact, it is not clear whether tensor programs can achieve good performance, beyond NNs.}  %
    \item[C3:\!\!]\emph{Data Representation}. %
    To use TCRs as execution engines, relational tables must be transformed into a tensor representation. Previous approaches have explored this challenge (e.g.,~\cite{tcudb}), but their cost of translation is not negligible. 
    Furthermore, TCRs commonly do not support strings or date data types.
    \item[C4:\!\!]\emph{Extensibility}. Running relational queries over TCRs makes running a query seamlessly over different hardware (CPU, GPU, ASICs, etc.) and backends (single node, distributed, edge, web browser, etc.) possible. 
    A single monolithic compiler architecture does not work in all situations, therefore \system's design must be flexible enough to address all these use cases.
\end{description}

\vspace{-2ex}
\subsection{Design Choices}
\label{sec:design}

When building \system, %
we embrace the following design choices. %
\begin{description}[style=multiline,wide,nosep]
    \item[DC1:\!\!] \emph{Avoid implementing data-dependent control flow in Python}. As Table~\ref{tbl:filter_example} suggests, computation in \system must use tensor operations as much as possible. Note that for loops and conditionals over schema elements are acceptable (e.g., loops over the columns of a table). This design choice allows us to address \textbf{C2} and achieve \textbf{G1}.
    \item[DC2:\!\!]  \emph{\revision{Tensor-based} columnar format for input tabular data.} Relational data must be transformed into the tensor format. To do this, \system adopts a columnar representation of tables, and considers each column in a table as a tensor. We provide more details on our data representation in \cref{sec:datamod}.
    This design choice addresses \textbf{C3}. %
    \item[DC3:\!\!]  \emph{Adherence to TCRs' API.}
    This design choice is required for achieving \textbf{G2} and \textbf{G3}. 
    In fact, if we start extending TCRs with new features and operators, eventually the system will hinter portability and increase the engineering effort because we will have to support them on any hardware. 
    Hence, we take advantage of existing TCRs' API rather than try to extend them. 
    \revision{With this design choice, we are also able to address \textbf{C1}}.
    \item[DC4:\!\!]  \emph{Extensible infrastructure allowing easy integration with relational and ML frameworks.}
    Having a flexible infrastructure is of paramount importance since we desire to ride the wave of investments in ML. Therefore, we embrace an extensible architecture that allows different output target formats (e.g., PyTorch, ONNX), composed of a core compiler, pluggable frontends (e.g., query parser and optimizer). This design choice addresses \textbf{C4}.
    
\end{description}

%% file: system.tex
\vspace{-1ex}
\section{Tensor Query Processor (\system)}
\label{sec:surakav}

In \system, relational operators and ML models are compiled into tensor programs using a unified infrastructure, extended from {\sc Hummingbird}~\cite{hummingbird,hummingbird-github}. 
Here, we focus on the relational operator part, as the ML part was described in~\cite{hummingbird}.

\myparagraph{\system Overview.} \system's workflow has two phases: (1) \emph{compilation}: an input query is transformed into an executable tensor program; (2) \emph{execution}: input data is first transformed into tensors, and then fed into the compiled program to generate the query result. 
Currently, \system uses vanilla PyTorch 
in the compilation phase as the implementation target for the tensor programs. 
If necessary, PyTorch programs are lowered into different target formats for portability or performance goals.
The selection of the hardware device to target is generally made in the compilation phase. %
Next, we first describe how \system represents relational data using tensors (\cref{sec:datamod}), and then describe each phase in detail (\cref{sec:compilation} and \cref{sec:execution}).

\dataRepFigure

\subsection{Data Representation}
\label{sec:datamod}

Before executing the query, \system must convert the input (tabular) data to tensors. 
\revision{Databases often manage and convert
data into their own proprietary format, and \system is no different.}
\system internally represents tabular data in a columnar format with virtual IDs~\cite{column-stores}, as illustrated in Figure~\ref{fig:data_rep}. 
Data for each column is stored as a $(n \times m)$ tensor, where $n$ is the input number of rows, and $m$ is the length required to store the values. 
The translation logic is different depending on the column data type. 
For example, \textit{numerical} columns ({\sf sid} in \cref{fig:data_rep}) can be directly represented as $(n \times 1)$ tensors. 
The conversion of numerical columns to tensors is often zero-copy. 
\system represents \textit{date} data in $(n \times 1)$ numeric tensors as the number of nanoseconds since some pre-defined epoch. 
In this case, (de)serialization may be required depending on the source/target \textit{date} representation. 
Finally, \system represents string columns using $(n \times m)$ numeric tensors, where $m$ is the maximum character length of any string for that column. 
Given a string, \system stores a character per tensor column and right-pads it with $0$s if its length is smaller than $m$. %
\revision{We are actively working on adding support for encoded data (e.g., bit packing, run-length encoding, dictionary encoding) and more compact string representations~\cite{nvidia-blog-string}.}

\vspace{-1ex}
\subsection{Query Compilation}
\label{sec:compilation}

\system's compilation phase is composed of four main layers, as shown in \cref{fig:systemFigure}: (1)~The Parsing Layer (\cref{sec:parsing}) converts an input SQL statement into an internal \emph{intermediate representation (IR)} graph depicting the query's physical plan, which is generated by an external \emph{frontend} database system. 
The architecture decouples the physical plan specification from the other layers, therefore allowing to plug different frontends. 
(2)~The Canonicalization and Optimization Layer (\cref{sec:optimization}) performs IR-to-IR transformations. (3)~The Planning Layer (\cref{sec:planning}) translates the IR graph generated in the previous layer into an \emph{operator plan} in which each operator is mapped into a tensor program implementation. 
(4)~The Execution Layer (\cref{sec:executionl}), using the operator plan, generates an \emph{executor} which is the program that \revision{runs} on the target TCR and hardware. 
Next, before describing each layer in more detail, we give a quick overview of \system's intermediate representation (IR).

\systemFigure

\subsubsection{Intermediate Representation (IR)}
The IR is a graph-based data structure. 
It consists of a list of \textit{operators} and \textit{variables}. %
Each \emph{operator} corresponds to a node in the graph, and it contains: (1)~a list of input variables; (2)~a list of output variables; (3)~an \textit{alias} identifying the operator type;  %
and (4)~a \emph{reference} to the corresponding operator instance in the original physical plan. 
The latter is used to instantiate the tensor program implementing the operator. For example, to create a filter, \system needs to access the expressions contained in the original physical operator.%

Edges represent data (tensors) flowing between operators. In particular, an edge connects an output variable from an operator to an input variable of another operator. 
A \emph{variable} contains: (1)~a unique identifier, and (2)~the corresponding frontend column name in the original plan, which is used to translate expressions. 
When a variable is created, a unique identifier is generated deterministically based on information available in the graph. Variables in the IR are generated as follows. 
First, \system generates a variable for each column in the input table. 
Then, these variables can be used as input to many operators; however, a new variable will always be created for an output of an operator. 
Thanks to this design: (1)~properties (e.g., sorting information) can be immutably attached to columns; (2)~the IR is easier to debug because variables, once defined, are never changed; and (3)~\system can detect at runtime when a column is not used anymore and safely garbage-collect it.

\eat{The input and output variables of an operator, which represent its input and output columns, form its \emph{schema}. %
When the IR graph is iterated over, the schema information of each operator allows to reconstruct the dependency chain among the operators. This is lazily computed every time the IR graph must be traversed. %
}

\subsubsection{Parsing Layer}
\label{sec:parsing}

The goal of the Parsing Layer is to translate input queries into \system's internal IR.
This goal is accomplished in two steps: (1)~input queries are parsed, optimized, and exposed as frontend-specific physical query plans; and (2)~a frontend-specific parsing logic translates the physical plan into an IR plan.

In its current version, \system supports queries expressed as Spark SQL statements, and it uses the PySpark API to parse, optimize, and return the physical plan in a JSON format. 
\revision{We plan to add support for Calcite~\cite{calcite}, DuckDB~\cite{duckdb}, and eventually Substrait~\cite{substrait}\footnote{Note that we currently only support Apache Spark for relational frontends, not in general. 
\system, in fact, supports all the ML frontends available in {\sc Hummingbird~\cite{hummingbird-github}}.}.}
Then the Spark parser constructs the internal IR version of the physical plan using a DFS post-order traversal. 
If an unsupported operator is found in the plan, this phase will fail with an exception.
The list of operators supported by the IR is extensible (\textbf{DC4}). %

\subsubsection{Canonicalization and Optimization Layer}
\label{sec:optimization}

This layer implements IR graph transformations similarly to a classical rule-based optimizer. 
Rules are applied to the IR graph in two stages. In the first stage, \emph{canonicalization}, the rules are used to eliminate any of the frontend-system idiosyncrasies in the IR graph. 
For example, Apache Spark returns a projection operator with no inputs for {\sc count *} statements.
In the second stage, \emph{optimization}, rules rewrite the IR graph for obtaining better performance. 
While we did not explore in depth the optimization space enabled by \system's design, we show that hand-optimized tensor programs are more efficient than the one currently generated by \system in \cref{sec:ex-manual}. 

\subsubsection {Planning Layer}
\label{sec:planning}

In this layer, \system transforms the optimized IR graph into an operator plan composed of PyTorch tensor programs implementing each operator in the IR graph. 
In \cref{sec:algorithms}, we describe some operator implementations in detail. 
The implementation of the Planning Layer is straightforward. 
For each operator in the IR graph, \system fetches the corresponding implementation containing the tensor program from a dictionary, which is then instantiated with the IR operator's reference to the frontend physical operator instance.

\subsubsection{Execution Layer}
\label{sec:executionl}

Here the operator plan is wrapped around a PyTorch \textit{executor} object. This object is responsible for: (1)~calling the tensor programs in the operator plan following a topological order; (2)~wiring the output tensors generated by each program into the successive one; and (3)~keeping track of tensor references to garbage collect them if not used anymore. 
Once the executor program is generated, \system provides options to compile it into different \emph{target formats} in addition to PyTorch interpreted execution. 
Currently, \system allows lowering the executor into the TorchScript and ONNX formats, as well as to use TVM to compile it directly into machine-level code. 
Note that not all queries can be compiled into all formats since not all tensor operations are supported by all the target formats.

\subsection{Execution}
\label{sec:execution}

Once the executor program is generated, 
it can be executed over the input data. 
The program automatically manages (1)~converting data into the tensor format;  %
(2)~data movements to/from device memory; and (3)~scheduling of the operators in the selected device.
Once the data is in the proper format and on the desired device, all the operators are executed sequentially. 
Regarding parallelization, \system exploits the tensor-level intra-operator %
parallelism provided by the TCRs. 
However, given the poor scalability performance (\cref{sec:ex-scaling}), we are exploring support for inter-operator parallelism and data-parallel strategies. 
Once the executor completes, %
\system returns the query result in tensor, NumPy, or Pandas formats.

%% file: compilation.tex
\section{Operator Implementation In TQP}
\label{sec:algorithms}

We described how \system uses the Planning Layer to translate relational operators in the IR graph into tensor programs. 
Here we provide an overview of a few program implementations.
\system provides tensor-based implementations for the following relational operators: selection, projection, sort, group-by aggregation (sort-based), natural join (hash-based and sort-based), \revision{non-equi}, left-outer, left-semi, and left-anti joins. 
\system supports expressions including comparison and arithmetic operations, functions on \textit{date} data type, {\sc in}, {\sc case}, {\sc like} statements, as well as aggregate expressions using {\sc sum}, {\sc avg}, {\sc min}, {\sc max}, and {\sc count} aggregates (with and without {\sc distinct}).
Finally, \revision{\system supports \textit{nulls}, and subqueries (scalar, nested, and correlated), and {\sc predict} UDF\footnote{\revision{While generic UDFs are hard to support in \system because of data conversion and data representation mismatches, Spark vectorized UDFs~\cite{vectorized-udf} can be %
supported on CPU.}}~\cite{predictsql,predict-spark-synapse}.}
With all the above, \system is able to compile and execute all 22 queries of the TPC-H benchmark (\textbf{C1}). 
Interestingly, to support the full TPC-H benchmark, only the tensor operations listed in \cref{sec:tensor-operations} are required, and we did not have to introduce any additional custom tensor operators (\textbf{DC3}). 
Due to space constraints, 
we only describe how \system implements relational expressions with tensor operations (\cref{sec:expressions}), and implementations for two representative operators: \revision{join (sort- and hash-based, in \cref{sec:sort-join} and \cref{sec:hash-join}, respectively)}, and group-by aggregation (\cref{sec:aggregate}). 
\revision{Finally, note that the filter implementation in \system is close to the Bitmap representation described in \cref{sec:tensor-program-example}.} 

\setlength{\textfloatsep}{0pt}

\subsection{Expressions}
\label{sec:expressions}

Relational expressions such as {\sc sum(l\_extendedprice $*$ (1 - l\_discount))} can be found in projection operators, filter conditions, etc. 
In an expression tree, each leaf node represents a column or a constant value (e.g., {\sc l\_extendedprice}) and each branch node represents an operator (e.g., {\sc $*$}).
\system keeps an internal dictionary that maps operators to their corresponding tensor operations, e.g., {\sc $*$} to \texttt{torch.mul}. 
To implement an expression with tensor operations, \system then performs a post-order DFS traversal on the expression tree. 
For each leaf node, \system fetches (or generates) the proper column-tensor (constant value). 
For each internal operator, \system retrieves the corresponding tensor operation (or a series of tensor operations) from the internal dictionary. 
In this way (and with the help of Python lambda functions), \system generates a chain of tensor operations representing the evaluation of the expressions.
As an example, from Q21 in TPC-H, the expressions {\sc o\_orderstatus = `f' and receiptdate > l\_commitdate} is implemented as \texttt{torch.logical\_and(torch.eq(o\_orderstatus,[70])
,torch.gt(l\_receiptdate,l\_commitdate))}, where \texttt{[70]} is a 1x1 tensor storing the ASCII value for the constant `\texttt{F}'.

\subsection{Sort-Based Join}
\label{sec:sort-join}

\GenericSortMergeJoinShort
\joinIllustration

\revision{\system adopts a late materialization strategy for joins, similar to the one commonly used in columnar databases~\cite{cstore-book,jive-join}. \system takes only the columns in the join predicate as input to the join, and the output is a set of pairs of indexes identifying the records for which the join predicate succeeds.} 
The sort-based equi-join algorithm is shown in \cref{alg:GenericSortMergeJoinShort}, where, to simplify the description, we describe the case in which two integer columns are joined. 
\revision{With a few modifications, the algorithm is also able to support non-equi joins, left-semi joins, and outer joins.}
We use the typewriter font (e.g., \texttt{bucketize}) to denote tensor operations, and the capital font (e.g., {\sc createOutput}) to denote class methods. 
\cref{fig:joinIllustration} further illustrates the algorithm.

First, \system sorts the join-key columns from each table (lines $1$ to $3$ in \cref{alg:GenericSortMergeJoinShort}, ~\ding{202} in \cref{fig:joinIllustration}). %
Then,~\ding{203}, \system builds two histograms for the join keys from \textit{left} and \textit{right}, respectively, i.e., \system counts the number of occurrences for each unique join key (line $4$). 
Then,~\ding{204} by multiplying the values (element-wise) of the histograms (line $5$), \system computes the bucket sizes: the number of output rows for each matching join key from \textit{left} and \textit{right}. 
Afterward, \system computes the prefix sums for the \textit{left} and \textit{right} histograms (\ding{205}), as well as their element-wise multiplication (\ding{206}) (lines $6$ to $8$). 
The prefix sums will be used later to retrieve, from each join output, the position in \textit{left} and \textit{right}.
The total size of the output of the join is then computed as the last element of the prefix sum containing the bucket sizes (line $9$), and ~\ding{207} \system generates an index array (\emph{offset}) of the same size (line $10$).
Then,~\ding{208} \system performs a parallel binary search on the prefix sum containing the bucket sizes to find the matching join key (bucket) to which each row in the output of the join belongs (line $11$). 
Next,~\ding{209} \system computes the indexes from \textit{left} and \textit{right} that generate each row in the output of the join. 
\cref{fig:joinIllustration} shows the computation process for row $8$ in the join output of the example. 
To compute the indexes from \textit{left} and \textit{right} that are part of a given offset in the output of the join, \system first subtracts \emph{offset} by the prefix sum of bucket sizes prior to the current bucket (line $12$). 
Now \emph{offset} becomes the offset in each bucket of the matching join keys. 
\system then adds to the \emph{offset} the previous bucket from the respective prefix sum histogram (\emph{cumLeftHist} and \emph{cumRightHist}, respectively), and adds the result (quotient for \emph{leftOutIdx}, remainder for \emph{rightOutIdx}) of \emph{offset} divided by the number of join keys from \textit{right} in the current bucket of matching join keys (lines $13$ to $14$). 
Finally, for each row in the join output, \system knows which rows from \textit{left} and \textit{right} contributed to it. 
It then generates the join output (line $15$, not depicted in \cref{fig:joinIllustration}).
It is important to note that all computations in this join implementation are achieved using tensor operations, with only minimal usage of Python code.

\setlength{\textfloatsep}{0pt}
\begin{algorithm}[t!]
    \small
    \color{\revisioncolor}
    \hspace{-10ex}\textbf{Input:} {\normalfont \emph{data}: input columns passed as an array of tensors.}
        
    \hspace{-10ex}\textbf{Output:} {\normalfont an array of tensors representing the join output.}
        \begin{algorithmic}[1]

                \State $\mathit{left}, \mathit{right} \gets \Call{getJoinKeyColumns}{\mathit{data}}$
                
                \State $\mathit{leftIdx}, \mathit{rightIdx} \gets \texttt{arange}(\mathit{left.shape[0]}),  \texttt{arange}(\mathit{right.shape[0]})$
                
                \hspace{-7ex}$\triangleright$  Compute the hash values for join keys (m is the max hash table size)
                \State $\mathit{leftHash}, \mathit{rightHash} \gets \texttt{remainder}(\mathit{left}, \mathit{m}), \texttt{remainder}(\mathit{right}, \mathit{m})$

                \hspace{-7ex}$\triangleright$  Build the histogram of hash values for the left join keys
                \State $\mathit{hashBincount} \gets \texttt{bincount}(\mathit{leftHash})$
                \State $\mathit{maxHashBucketSize} \gets \texttt{max}(\mathit{hashBincount})$

                \hspace{-7ex}$\triangleright$  Build and probe the hash table in an interleaved way
                \For {$\mathit{i} \in range(\mathit{maxHashBucketSize})$}
                    
                    \State $\mathit{hashTable} \gets \texttt{full}((\mathit{m} + 1, ), -1)$
                    
                    \State $\mathit{hashTable}.\texttt{scatter\_}(0, \mathit{leftHash}, \mathit{leftIdx})$

                    \hspace{-7ex}$\triangleright$  Skip those scattered for future iterations by setting their hashes to $\mathit{m}$
                    \State $\mathit{leftIdxSct} \gets \texttt{masked\_select}(\mathit{hashTable}, \mathit{hashTable} \ge 0)$
                    \State $\mathit{leftHash[leftIdxSct]} \gets \mathit{m}$
                    
                    \hspace{-7ex}$\triangleright$  Probe the current hash table and get the left and right indexes
                    \State $\mathit{leftCandIdx} \gets \mathit{hashTable[rightHash]}$
                    
                    \State $\mathit{validKeyMask} \gets \mathit{leftCandIdx} \ge 0$
                    
                    \State $\mathit{validLeftIdx} \gets \texttt{masked\_select}(\mathit{leftCandIdx}, \mathit{validKeyMask})$
                    
                    \State $\mathit{validRightIdx} \gets \texttt{masked\_select}(\mathit{rightIdx}, \mathit{validKeyMask})$
                    
                    \hspace{-7ex}$\triangleright$  Find the indexes that have matching join keys
                    
                    \State $\mathit{matchMask} \gets \mathit{left[validLeftIdx]} == \mathit{right[validRightIdx]}$
                    \State $\mathit{leftMatchIdx} \gets \texttt{masked\_select}(\mathit{validleftIdx}, \mathit{matchMask})$
                    \State $\mathit{rightMatchIdx} \gets \texttt{masked\_select}(\mathit{validrightIdx}, \mathit{matchMask})$
                    
                    \hspace{-7ex}$\triangleright$  Append the indexes to the global results
                    \State $\mathit{leftOutIdx} \gets \texttt{cat}((\mathit{leftOutIdx}, \mathit{leftMatchIdx}))$
                    \State $\mathit{rightOutIdx} \gets \texttt{cat}((\mathit{rightOutIdx}, \mathit{rightMatchIdx}))$
                \EndFor
                
                \State $\Return \text{ } \Call{createOutput}{\mathit{data}, \mathit{leftOutIdx}, \mathit{rightOutIdx}}$

        \end{algorithmic}

        \caption{Hash-Based Join}

        \label{alg:GenericHashJoinShort}
\end{algorithm}

\revision{
\vspace{-2ex}
\subsection{Hash-Based Join}\label{sec:hash-join}
The hash equi-join algorithm is shown in \cref{alg:GenericHashJoinShort}. 
The definition of the input and output here is the same as in \cref{sec:sort-join}. 
The algorithm is similar to the classical hash join algorithm, except that the build and probe phases are interleaved and repeated as many times as the maximum number of elements that share a hash value (line 6). 
The algorithm is as follows: 
\system first generates the indexes (line 2) and the hash values (line 3) for the $\mathit{left}$ and $\mathit{right}$ tables.
Afterward, \system computes a histogram over the table on which the hash table will be built ($\mathit{left}$ in this case, line 4) and checks the maximum number of elements in a hash bucket (line 5).
Then, \system repeatedly builds a hash table (lines 7 and 8) and probes it (lines 11 to 14) to find matching keys (lines 15 to 17).
Matching keys are accumulated across iterations (lines 18 and 19). 
In each iteration, \system also keeps track of the indexes that are stored in the hash table such that they will not appear in subsequent iterations (lines 9 and 10). 
To achieve this, let $m$ be the hash table size; \system appends an additional $\mathit{(m+1)}$-th bucket to the hash table and uses it to redirect the already scattered indexes. 
Note that when there are no hash collisions, \system skips the logic of lines 9 to 10 and 18 to 19. 
This path is therefore close to the optimal.}

\revision{Compared to the sort-based join, when there are no hash collisions, this implementation is around 30\% to 50\% faster on CPU and 2$\times$ faster on GPU. 
When there are hash collisions, it is faster than the sort-based join for cases in which at most around 15 elements share a hash value; when there are more than 15 elements sharing a hash value, the sort-based join is faster. We are currently working on a partitioned hash-join implementation.
}

\vspace{-1ex}
\subsection{Aggregation}
\label{sec:aggregate}
\setlength{\textfloatsep}{8pt}
\AggregationShort

\cref{alg:AggregationShort} shows the pseudocode of the aggregation implementation. 
First, \system horizontally concatenates the values of the group-by columns (lines $1$ and $2$). 
\system then sorts the values of the concatenated columns using radix sort and permutes all the input data columns according to this sorted order (lines $3$ and $4$). 
Using \texttt{uniqueConsecutive}, \system eliminates all but the first key from every consecutive group of equivalent keys. 
Concurrently, \system computes the inverted indexes that indicate which bucket (unique key) each row in the sorted list ends up in (line $5$). 
Finally, with the unique key list and inverted indexes, \system evaluates the aggregate expression for all groups. 
This last operation makes use of the expression generated (at compile time) as described in \cref{sec:expressions}.

%% file: experiments.tex
\vspace{-1.5ex}
\section{Evaluation}
\label{sec:experiments}
\vspace{-0.2ex}

\revision{
The evaluation aims to answer the following questions: 
(1) On CPU, is \system's performance comparable to other data processing systems on a single core (\cref{sec:ex-single-core})? (2) On GPU, is \system's performance comparable to other GPU databases (\cref{sec:ex-gpu})? (3) How well does \system scale with the increase in the number of CPU cores and dataset sizes (\cref{sec:ex-scaling})? (4) What is the cost/performance trade-off of \system on GPU (\cref{sec:cost-performance})? (5) Which operation takes the most time in query execution (\cref{sec:ex-breakdown})? (6) Can hand-optimized query plans improve \system's query time (\cref{sec:ex-manual})? (7) Can \system accelerate workloads mixing ML and relational queries (\cref{sec:ex-predictive})? (8) What are the overheads (\cref{sec:ex-overheads})? (9) Can \system run over different hardware and software backends while minimizing the engineering effort (\cref{sec:ex-portability} and \cref{sec:ex-engineering})?

}

\stitle{Baseline systems.}
Our goal is to compare \system with state-of-the-art query processing systems for different hardware settings.
Specifically, for CPU execution, we compare \system with Apache Spark~\cite{spark} (recall that Spark and \system share the same query plans) and DuckDB~\cite{duckdb}: a \revision{state-of-the-art} vectorized engine. 
For GPU execution, we compare \system with two well-known open-source GPU databases: BlazingSQL~\cite{blazing-sql} and OmnisciDB~\cite{omniscidb}. 

\stitle{Hardware and software setup.} For all the experiments (except when noted otherwise), we use an Azure NC6 v2 machine with 112 GB of RAM, an Intel Xeon CPU E5-2690 v4 @ 2.6GHz (6 virtual cores), and an NVIDIA P100 GPU (with 16 GB of memory). 
The machine runs Ubuntu 18.04 with PyTorch 1.1\revision{1}, torch-scatter 2.0.9, BlazingSQL 21.8.1, PySpark 3.1.1, OmnisciDB 5.9.0, DuckDB 0.4.0, RAPIDS 21.08, CUDA 10.2, TVM 0.8 and scikit-learn 0.21.3.

\stitle{Experimental setup.}
We use the TPC-H benchmark~\cite{tpc} which consists of 22 queries. 
We use the parameters specified in the query validation sections in \cite{tpc}. 
We generate data at different scale factors (from 1 to 10 where 1 means 1 GB of data in total\revision{\footnote{\revision{Note that some queries can run on scale factors larger than 10 in GPUs, thanks to
TQP’s ability to push projections into data conversion. We are working on
supporting out-of-memory computation by leveraging PyTorch’s DataLoader~\cite{dataloader}.}}})
using the dbgen tool. 
We load the generated data from disk into Pandas dataframes. \revision{All dataframes use the data types as specified in the benchmark, except for decimals: we use doubles for all systems since \system does not support decimals yet.}
Subsequently, we register/convert each dataframe into each system's internal format, e.g., Spark dataframes for Spark\footnote{For Spark, we additionally load the working datasets in memory using \texttt{cache}.}, PyTorch tensors for \system, CUDA dataframes for BlazingSQL, etc., and move the data to the GPU, when applicable.
We measure the total query execution time, including the time for generating the output. 
For each experiment, we do 10 runs where the first 5 are for warm-up. 
The reported numbers are median values of the last 5 runs. 

\stitle{Key takeaways.}
(1) \system's query execution time on CPU using a single core is better than Spark's over the same physical plans; 
however, (2) \system's scalability on CPU is poor because of PyTorch lacking parallelization in some operators' implementation and its intra-operator parallelism model. 
\revision{(3) \system is, in general, slower than DuckDB on CPU, but for a few queries, \system is comparable or even better.} 
(4) Hand-optimized plans can improve \system's performance, which suggests that a TCR-aware query optimizer is required to achieve the best performance. 
(5) \system's query execution time on GPU is usually \revision{better than both BlazingSQL's and OmnisciDB's}, and \system supports more queries than they do. 
(6) When ML model prediction and SQL queries are mixed together, \system is able to provide end-to-end acceleration which delivers up to \revision{$9\times$} performance improvement over CPU baselines. 
\revision{(7) \system on GPU performs favorably, and the query time speedup justifies the dollar cost increase compared to CPU-only systems. 
(8)} \system can run queries on different hardware and software backends (including even integrated GPUs and web browsers), with orders of magnitude fewer lines of code required compared to the baseline systems.

\setlength{\tabcolsep}{3pt}
\evalCpuGpuTableSmall
\setlength{\tabcolsep}{6pt}
\evalScalabilityFigure
\vspace{-1ex}
\subsection{Single Core Execution on CPU}
\label{sec:ex-single-core}

In this first experiment, we use a single CPU core and TPC-H at scale factor 1. 
The results are shown in \cref{tbl:evalCpuGpuTable} (under CPU). 
We compare Spark and DuckDB vs. \system, using both interpreted (\system) and compiled execution with TorchScript \revision{(TQPJ)}. 
Spark, DuckDB, and \system can support all 22 queries. 

In terms of query time, TQPJ is either comparable to \system or better. 
This is because TorchScript removes Python code dependency and provides optimizations \revision{not offered by vanilla PyTorch}~\cite{torchscript-optimizations}. 
\system outperforms Spark for most queries, sometimes by an order of magnitude (e.g., \revision{Q10, Q15, and Q22}). 
Given that \system uses the same physical plans as Spark, this suggests that the tensor abstraction is indeed good for executing relational queries. 
The practical reasons are: (1) \system is column-oriented, while Spark is row-oriented. This makes the former better suited for analytical queries; (2) some tensor operations use SIMD instructions, while Spark does not exploit vectorization; (3) in \system, tensor operations are implemented in C++, while Spark is Java-based; \revision{(4) Spark is designed as a scale-out system.}
For queries (i.e., Q1, Q13, and Q21) where \system is slower than Spark, the reasons are: (1) \system's left anti-join and left outer-join implementations are not optimized; 
(2) the performance of the \texttt{uniqueConsecutive} operator in PyTorch is not optimal. 
Finally, %
\system has better performance than DuckDB only for 3 queries. 
For the other queries, DuckDB clearly outperforms \system. 
If we exclude Q1, Q13, and Q21 (discussed above), \system's query times are within the same order of magnitude as DuckDB's. 
To evaluate whether this poor performance compared with DuckDB is due to bad query plans or the tensor abstraction, we hand-code better query plans and tensor programs in \cref{sec:ex-manual} and show that \system can match and even outperform DuckDB on CPU.

\vspace{-1ex}
\subsection{Execution on GPU}
\label{sec:ex-gpu}
In this experiment, we evaluate the performance of \system on GPU. 
The results are shown in \cref{tbl:evalCpuGpuTable} (under GPU). 
Starting from \system vs. TQPJ, \revision{as in the CPU case, TQPJ outperforms \system.}
\revision{
Compared with the baselines, \system (interpreted or compiled) outperforms BlazingSQL \revision{(Blazing in the table)} for all the queries, and it outperforms OmnisciDB \revision{(Omnisci)} on 15 queries out of the 18 queries supported by OmnisciDB. 
For the remaining 3 queries, \system achieves query times within a factor of 2 from OmnisciDB. Note that \system supports all 22 TPC-H queries, while BlazingSQL and OmnisciDB only support 17 and 18 queries, respectively.}

Finally, if we compare the best CPU performance versus the best GPU ones, in general, we see that the query times on GPU are 1.5$\times$ to \revision{48$\times$} better than the CPU ones (single core), except for Q16 where DuckDB is about \revision{3$\times$} faster than the best-performing GPU system. 
This somehow counter-intuitive result is due to the fact that, at scale factor 1, GPU resources are not completely saturated. Therefore, it makes sense to explore how these systems scale with more data and more available core. 
This is what we explore next.

\vspace{-1ex}
\subsection{Scalability}
\label{sec:ex-scaling}

For this and the following experiments, we select a representative set of queries: complex aggregation (Q1), joins and filters (Q2), simple filters (Q6), complex joins (Q9), simple join and aggregation (Q14), a complex mix of join, aggregation, and sub-queries (Q18). %
\vspace{-1ex}
\subsubsection{Scaling the Number of Cores}
In this experiment, we scale the number of available CPU cores from 1 to 6 over TPC-H at scale factor 1.
Figure~\ref{fig:evalScaleThreadFigure} compares the scaling performance of Spark, DuckDB, and \system. 
Spark has the best scalability trend lines almost for all queries. 
DuckDB also scales well. %
\system's scaling performance is, however sub-optimal, and for some queries increasing the number of cores provides no benefits. %
There are two reasons: (1) PyTorch uses \revision{intra-operator} parallelism %
, which is not as efficient as the shuffle~\cite{spark} or morsel-based~\cite{morsel} approaches in Spark and DuckDB, respectively; (2) some PyTorch operators run on a single core (e.g., \texttt{unique} and \texttt{unique\_consecutive}~\cite{torch-unique} used in aggregation).
\revision{We are investigating how to overcome this limitation by adding data-parallel support to \system leveraging PyTorch Distributed Data Parallel~\cite{distributed-pytorch,ddp} or by adding parallel operator implementations.}

\subsubsection{Scaling the Data}

In this experiment, we scale the dataset from 1 GB to 10 GB. 
In \cref{fig:evalScaleDatasetFigure}, we compare the scalability performance of CPU implementations running over 6 cores (Spark, DuckDB), as well as GPU systems (BlazingSQL and OmnisciDB). 
In general, we see that \system CPU scales the worst for almost all queries (only Spark is worst for Q6 and Q14), while GPU systems scale better than the CPU ones.
For Q1, OmnisciDB provides the best performance, followed by \system GPU. 
For Q2, Q14, and Q18, \system GPU has the best performance, while for Q6, \system GPU is comparable to OmnisciDB. 
Finally, for Q9, OmnisciDB has the best performance. 
Q9 has six joins, and OmnisciDB is able to better use the GPU resources. 
This query is memory-bound, and the memory bandwidth of the P100 makes it much faster on GPU than on CPU. %

\evalCostPerformanceFigure
\evalOpsBreakdownFigure
\evalUtilizationFigure
\revision{
\vspace{-1.5ex}
\subsection{Cost/Performance Trade-off}
\label{sec:cost-performance}
\vspace{-0.5ex}

We now provide a cost/performance analysis of \system on GPU compared to a CPU-only baseline. 
Specifically, we select a general-purpose (CPU-only) VM in Azure with a dollar cost similar to the cheapest VM equipped with GPU (NC4as\_T4\_v3), and with similar main memory size. Following these constraints, we select a D2ds\_v5 with 8 CPU cores and 32GB of memory.
Then we compare the performance of DuckDB on the D2ds\_v5 with \system on (1) NC4as\_T4\_v3 (with an NVIDIA T4 GPU, about 15\% more expensive than the CPU-only machine), (2) NC6s\_v2 (with an NVIDIA P100, around 4.6$\times$ more expensive than the CPU-only VM), and (3) NC6s\_v3 (with an NVIDIA V100, around 6.6$\times$ more expensive than the CPU-only VM).
For each GPU VM type, we show the query time speedup required to be more cost-effective than the DuckDB baseline. 
That is, for the T4, the speedup provided by \system has to be more than $15\%$ to justify the cost increase of the T4 VM compared to the DuckDB CPU baseline, 4.6$\times$ for the P100, 6.6$\times$ for the V100. 
The results for scale factor 10 are shown in Figure~\ref{fig:cost-perf} for a few representative TPC-H queries. 
As shown, \system on GPU is more cost-effective compared to DuckDB on the CPU-only machine: for 6 of the 6 selected queries (17 of the 21 supported queries\footnote{OOM errors occurred when TQP ran Q21 at scale factor 10 on these GPUs.
} in the full TPC-H) for the T4; 5 of 6 (10 of 21 in the full TPC-H) for the P100; and 5 of 6 (9 of 21 in the full TPC-H) for the V100.
}

\vspace{-1ex}
\subsection{\revision{Performance Breakdown}}
\label{sec:ex-breakdown}

\revision{
In this experiment, we show the major contributing factors to the query execution time. 
\system is integrated with TensorBoard~\cite{tensorboard-github}, which provides performance breakdowns and makes it easy to spot bottlenecks~\cite{tqp-demo}.
We start by looking into which tensor operators are responsible for the majority of the execution time. Figures~\ref{fig:evalOpsBreakdownCpuFigure} and~\ref{fig:evalOpsBreakdownCudaFigure} show the breakdown for a few selected queries on CPU and GPU, respectively. 
Interestingly, even if \system uses the same algorithms on both CPU and GPU, the same query can show different operator contributions. 
For example, for Q1 on CPU, most of the time is spent on computing the unique elements, while on GPU, most is spent on {\sf scatter\_add}. 
This is because the quality of the operator implementations is different for CPU and GPU. 
Across queries, on CPU and GPU, the majority of time is also spent on different operators. 
On CPU, most queries are bounded by unique operators, {\sf masked\_select}, and indexing; on GPU, most of the time is spent on sorting, {\sf unique} and {\sf nonzero}. %
These observations suggest that: (1) the quality of kernels differs between CPU and GPU, e.g., after further investigation, we find that the GPU implementation of {\sf scatter\_add} is not optimal, 
and nonzero requires host/device synchronization~\cite{nonzero} (however, we believe that over time the community will fix such performance issues); and (2) it might be worth investigating backend-aware tensor algorithms.}

\revision{Finally, we report the GPU utilization for the same set of queries in Figure~\ref{fig:evalUtilizationFigure}. 
As we can see, each query has different utilization characteristics. For instance, Q1 contains complex aggregation, and it spends 87\% of the time on kernel execution; conversely, Q6 and Q14 are simple queries, and most of the time is spent allocating GPU memory. Finally, Q2 spends a considerable amount of time in generating the output on CPU.}

\vspace{-1ex}
\subsection{Hand-Optimized Plans}
\label{sec:ex-manual}

\evalHandCodedTable

Next, we study whether \system's performance can be improved with a better optimizer able to generate better tensor programs. 
\revision{To understand this, we hand-optimize the tensor programs for a few selected queries similarly to what a reasonable optimizer with knowledge about cardinalities and tensor characteristics would do, e.g., avoid sorting (or computing unique) over already sorted (or unique) columns, and select better join implementations.} 
The results are shown in Table~\ref{tbl:evalHandCodedTable}, where we report the best baseline for each setting (CPU 1 and 6 cores, and GPU), and over three execution modes: interpreted PyTorch (Torch), compiled TorchScript (JIT), and compiled using TVM. 
TVM only supports \revision{Q6 and Q14}. %

If we focus on the CPU numbers first, \system's performance is comparable to or even better than that of DuckDB's, while \system was much slower compared to DuckDB both on single- and multi-core execution when not using the hand-optimized plans.
\system is now faster than DuckDB for all queries over 1 CPU core, and two queries over 6 CPU cores. 
For some queries, \system is faster than DuckDB by a large margin, e.g., for Q6, 1-core TVM execution is 6$\times$ faster. 
This is because TVM uses code generation and operator fusion to minimize intermediate data materialization across operators. 
\revision{When scaling to 6 cores, \system scales well only for Q14, while DuckDB scales linearly.
For the other queries, \system's query times improve by at most 2$\times$.}
This again shows the limitations \revision{of PyTorch's scalability on CPU}%
, which cannot be improved by using better tensor programs.

Finally, on GPU, we see that OmnisciDB has still better performance for Q9, although \system's query time for Q9 on GPU improves by 4$\times$, \revision{when using the hand-optimized plans.}
\revision{This is because \system's aggregate implementation heavily uses sorting, while OmnisciDB uses hash-based implementations.} 

\vspace{-1ex}
\subsection{Prediction Queries}
\label{sec:ex-predictive}

We now investigate the performance benefits of using a \revision{unified} runtime for queries mixing relational and ML operators. 
We use \emph{prediction queries} as a use case, i.e., queries embedding a trained ML model performing predictions over some input data~\cite{predict-spark-synapse}. 
\revision{Recall that \system natively
 supports predictions of any PyTorch model (e.g., NNs), and traditional ML models through its integration with {\sc Hummingbird}.}
Here, we join the {\sc customer} and {\sc orders} tables in TPC-H (scale factor \revision{10}), and train a gradient boosting tree model (with 128 trees with max depths of 8) over a mix of categorical ({\sc c\_orderstatus}) and numerical features ({\sc c\_custkey, c\_nationkey, c\_acctbal, sum(o\_totalprice)}) after we apply one-hot encoding and feature scaling, respectively. 
We run a prediction query using the trained model over the query \revision{with two filter predicates added} ({\sc c\_mktsegment = `\textsc{building}’ and o\_orderdate >= date `\textsc{1993-10-01}’}).
Note that this prediction query mixes ML operators (tree ensemble, one-hot encoding, scaling, and concatenation) with relational ones (join, aggregation and filtering). We compare \system with two baselines: one where the prediction query is executed over Spark (MLlib~\cite{mllib} is used to build the model), and one where we use DuckDB for the relational part and scikit-learn~\cite{scikit} for the ML part\revision{\footnote{\revision{Note that moving data from DuckDB to scikit-learn is zero-copy since DuckDB can directly return data in Pandas dataframe format~\cite{duckdb-pandas}.}}}. 
Since \system subsumes {\sc Hummingbird}, it is able to compile both the ML and the relational operators of the query into a unified plan executable on TCRs. 
\cref{fig:evalPredictiveFigure} shows the result. 
For CPU single core, \system is about \revision{40\%} faster than Spark, while DuckDB+scikit-learn is about \revision{7$\times$} faster than \system.
When enabling all cores, Spark and DuckDB scale much better than \system, for the reasons described in \cref{sec:ex-scaling}. 
Finally, \system is able to exploit GPU acceleration end-to-end, which brings a \revision{9$\times$} improvement of query time %
compared to the best CPU baseline.

\evalPredictiveFigure

\evalOverheadFigure

\subsection{Overheads}
\label{sec:ex-overheads}

Next, we evaluate the overheads of \system for both CPU and GPU.
\revision{
The breakdown of the end-to-end execution with all overheads is shown in Figure~\ref{fig:evalOverhead}. 
Note that: (1) data conversion is done once and many databases (e.g., BlazingSQL, OmnisciDB, Spark, SQL Server, etc.) requires it; (2) \system pipelines data movement (to the GPU) with query execution (non-blocking IO), while for this experiment we explicitly make data movement blocking; (3) the machine in this experiment uses PCIe 3 which is $4\times$ slower than the latest version, PCIe 5; (4) query compilation can be cached, but here we report the full query compilation time as the sum of the time for the frontend database to generate the physical plan, and the time for \system to generate the final executable tensor program.}

\revision{If we focus first on the CPU side (Figure~\ref{fig:evalOverheadFigureCpu}), compilation and data conversion take the majority of the time only for simple queries (e.g., Q6), while for the other queries, the majority of the time is spent on the query execution. 
However, in the GPU case (Figure~\ref{fig:evalOverheadFigureGpu}), except for Q2 and Q9, the majority of the time is spent on data operations (conversion and movement) and compilation. 
However, in practice, as described above, these overheads are hidden (e.g., data movement using pipelining) or are one-time overheads (data conversion and query compilation). 
Regarding query compilation, 90\% of the time is spent initializing the PyTorch models from the Spark plans, and we are currently investigating how to speed up this process. 
Finally, using TorchScript adds substantial compilation overheads since queries are traced using input samples.
}

\eat{
\stitle{Query Compilation.} %
The total query compilation time is the sum of the time for the frontend to generate the physical plan, and the time for \system to generate the final executable. 
Compilation times are query-dependent and they run from hundreds of ms to tens of ms.
In the current implementation with Spark, the physical plan generation takes around $10\%$ of the time. 
GPU compilation time is almost in line with the CPU numbers. 
When TorchScript is the target format, the compilation time increases considerably because TorchScript requires to \emph{trace} the program, i.e., run some input through the program to register the execution trace. 
TorchScript compilation time can add from 2$\times$ to 10$\times$ more overheads. }

\evalPortabilityTable

\eat{
\stitle{Data Conversion.}
Data conversion time can go from adding less than 1\% over the execution time to over 70\%, with a median of 1.25\%. 
This is because data conversion time is strictly data type- and query-dependent. 
It is data-type dependent because numerical types are zero-copy, while string and date types require conversion to numerical types, which is more expensive. 
It is also query dependent because \system pushes projections into the data conversion step so that only the columns that are used in the query get converted. 
GPU execution does not require any more specific data conversion, although it adds some data movement costs. 

\stitle{GPU Data Movement.}
In general, moving data to the GPU is a costly operation because it has to go through the PCIe which has low bandwidth.
Data movement can add up to 55\% on top of the execution time, with a median of 19\%.%
}

\vspace{-1.5ex}
\subsection{Portability}
\label{sec:ex-portability}

To evaluate whether \system can run on different hardware and software backends, we run TPC-H Query 6 with the hand-optimized plan on: 
(1) two integrated graphic cards, one from Intel, and one from AMD; 
(2) two discrete GPUs from NVIDIA (K80 and V100: the former a generation before the P100 GPU used for the experiments in the previous sections; the latter one, one generation after); %
(3) a custom ASIC used for NN training and inference (TPU); 
and (4) a web browser. 
We use a scale factor of 1. The results are shown in Table~\ref{tbl:evalPortabilityTable}. 
This experiment proves the versatility of \system. 
For the integrated GPUs, we use TVM to code-generate the query using Metal~\cite{metal}. 
For the two discrete GPUs, we use vanilla PyTorch, while for the TPU, we use the XLA backend for PyTorch\footnote{Note that PyTorch/XLA does not support all the necessary tensor operations and the execution fallback to regular CPU for part of the query is not available.}~\cite{pytorch-xla}.
Finally, we are able to run the query in the browser by exporting it into the ONNX format and running it in Chrome using ONNX Runtime (ORT) for WebAssembly (WASM)~\cite{ort-wasm}.

\eat{We also evaluated other configurations: a machine with an AMD Radeon Instinct MI25, and browser execution with WebGL for GPU acceleration. 
For the first configuration, we tried both ROCm~\cite{rocm} and DirectML~\cite{directml}. 
Unfortunately, we are unable to properly set up ROCm\footnote{ROCm does not support Windows, while the GPU drivers for the Azure VM we tried were only available on Windows.}, while DirectML lacks support for some operators. 
For the second one, we tried ORT with WebGL, and found that it lacks support for some operators and CPU fallback is not yet available.}

\vspace{-2ex}
\subsection{Engineering Effort}
\label{sec:ex-engineering}

To demonstrate the minimal engineering effort required by \system to run queries over different hardware, we compare the lines of code for a few relational operators (hash and sort-based joins, aggregation) across all evaluated systems. 
For each relational operator and each system, we use cloc~\cite{adanial_cloc} to count the lines of source code (excluding comment and blank lines) from the files containing the algorithmic functionality of the operator. %
This is admittedly a subjective process, but we believe the numbers of lines of code can roughly reflect the engineering effort required to implement relational operators in each system. 
\cref{tbl:evalLOCTable} shows the results. 
Compared with the baselines, \system requires significantly lower engineering effort: up to 10$\times$ less compared to CPU implementations, and 50$\times$ less compared to GPU ones. 
It is worth noting that \system is able to target different hardware with the same implementation, so the engineering effort required for \system to scale over different hardware is constant. 
The other baseline systems do not share this property. 
For instance, to run Spark on GPU (e.g., using RAPIDS~\cite{spark-rapids}, the same backend of BlazingSQL), we would have to add the lines of code for the GPU implementation.

\evalLOCTable

%% file: related.tex
\vspace{-1ex}
\section{Related Work}
\label{sec:related}

\stitle{Common representation for relational and ML workloads.}
Since the '90s~\cite{sqlserver-data-mining}, there have been many works trying to integrate relational queries with data science and ML workloads~\cite{db-ml-tutorial,madlib,bismark,in-db-ml,ml2sql,spark-mllib,samsara,vertica-pred,raven,masq,bigquery,tidypredict,spores,sparse-relational,modin,polyframe,10.14778/3317315.3317323,10.14778/3457390.3457399,systemml,laradb,daphne,systemds,predictsql,slacid,redshift-ml}. 
To our knowledge, we are the first to propose executing relational queries over TCRs. 
Earlier attempts tried to run a few relational operators on the TPU using TensorFlow~\cite{query-tpu}. 
\system is orthogonal to previous efforts to optimize relational and tensor algebra (e.g., ~\cite{laradb,spores}), and we believe \system can leverage them to improve its performance further. 
An analysis of matrix query languages can be found in~\cite{matlang}. Here, we focus on TCRs' tensor interface, which is more flexible than a linear algebra API. %

SciDB~\cite{scidb, scidb_overview} is a database using arrays as the base data representation. 
TensorDB~\cite{tensordb} further proposes support for tensor data and decomposition operations inside databases.
SciDB, TensorDB, and \system suggest using a format closer to data science and ML to represent data. 
However, \system further exploits TCRs to run both relational and ML workloads on hardware accelerators.

\stitle{GPUs and hardware accelerators.}
Several systems have explored running relational queries over GPUs~\cite{crystal,rateupdb,yin-yang,pump-up,10.14778/3425879.3425890,10.1145/2882903.2915224,gpu-mainstream}. 
We refer readers to~\cite{gpu-database-survey} for a recent survey. 
However, the majority of them focus mostly on microbenchmarks, while, to our knowledge, only RateUpDB can support the full TPC-H benchmark. 
\system is able to run the TPC-H benchmark \revision{on both} CPU and GPU, thanks to TCRs' \revision{flexibility} to support different hardware backends. 
TCUDB~\cite{tcudb} suggests using the Tensor Core Unit (TCU) of GPUs for accelerating relational operators. 
TCUDB requires an expensive transformation from tables to matrices and also uses low-level CUDA kernels, while \system takes advantage of the high-level tensor interface of TCRs. 
GPUs are the default hardware for running neural network models. 
However, there has recently been a rise in custom ASICs~\cite{tpu,graphcore,cerebras,sambanova,neural-engine} purposely built for ML workloads. 
With \system, we propose a solution allowing us to run relational queries on any hardware supported by TCRs, since many ASICs~\cite{ipu-pytorch,cerebras-software,tpu} provide high-level interfaces directly through TCRs or are targetable through tensor compilers~\cite{tvm,mlir}.

\stitle{Query processing over heterogeneous hardware.}
Several recent works have started to explore query execution over heterogeneous hardware, such as CPU-GPU co-execution~\cite{10.1145/3485126,9556045,10.1007/s00778-018-0512-y,hetexchange,voodoo,ocelot,dandelion,heterogenous-pipelines}. 
Many of them rely on OpenCL~\cite{opencl} to target different hardware. 
However, targeting a common language (or similarly a generic compiler, e.g., MLIR~\cite{mlir}), requires non-trivial engineering effort since each device requires proper tuning~\cite{voodoo}, algorithms, and data structures (as well as abstractions/dialects in the MLIR case). 
Conversely, \system can natively run on any hardware supported by TCRs, and uses TCRs' tensor operation implementations and compilation stacks. 
Currently, the user has to specify which fragment of the query should run on which hardware, but we are exploring how to automate this and enable co-execution. 

A trend arises recently that suggests splitting relational operators into smaller functions that can be easily composed and efficiently dispatched over heterogeneous hardware ~\cite{sub-operators,deep-optimization,modularis}. 
\system fits in this trend, whereby tensor operations are sub-components.

\revision{\stitle{Vectorized execution, query compilation, and columnar databases.}
MonetDB/X100~\cite{monetdb} pioneered the vectorized execution model as well as the columnar data layout~\cite{cstore}.
\system follows a similar design, where data is stored in a columnar format with virtual IDs~\cite{cstore-book}, but each column is represented as a tensor.
Recent works, such as HyPer~\cite{hyper} and others~\cite{dblab,rof,umbra}, have focused on query compilation. 
Nevertheless, since (1) there is no clear winner between query compilation and vectorized execution~\cite{vectorized-vs-compiled} ; (2) many industry-grade systems use vectorized execution because it is easier to debug and profile~\cite{photon}; and (3) compiled systems start to move to vectorized execution (e.g., Spark with Photon), we evaluate \system against a state-of-the-art vectorized engine, DuckDB~\cite{duckdb}.

On the ML systems side, TensorFlow initially embraced a compiled (graph) execution~\cite{tensorflow2}, while PyTorch pioneered interpreted (eager) execution~\cite{pytorch}. %
Compilers~\cite{tvm,xla,torchscript,mlir,taco,cortex,cora} and optimization techniques~\cite{taso,rgs,janus} for neural networks are hot topics in the MLSys community. 
With \system, we aim to ride the wave of innovation in this domain. %
For TQP, interpreted vs. compiled execution is just another point in the query optimization space, since TCRs allow to switch between them seamlessly.
}

%% file: conclusion.tex
\vspace{-1.2ex}
\section{Conclusion}
\label{sec:conclusion}

We proposed \system, the first system able to run relational queries on TCRs. 
\system is able to take advantage of all the innovation poured into TCRs, as well as to run efficiently on any hardware devices supported by TCRs. Our experiments showed not only that \system is capable of running the full TPC-H benchmark on TCRs, but also that \system's performance is comparable and often superior to that of specialized CPU and GPU query processing systems.

\vspace{-1ex} 
\begin{acks}
We would like to thank Yuki Asada, Victor Fu, Apurva Gandhi, Lihao Zhang, Advitya Gemawat, Venkatesh Emani, Masahiro Masuda, Ziheng Jiang, Raghu Ramakrishnan, and Magdalena Balazinska for their insightful feedback and support.
\end{acks}